\documentclass[12pt]{article}
\usepackage{epsfig}

\newcommand{\be}{\begin{equation}}
\newcommand{\ee}{\end{equation}}
\newcommand{\bea}{\begin{eqnarray}}
\newcommand{\eea}{\end{eqnarray}}

\newcommand{\gta}{\mathrel{\hbox to 0pt{\lower 3pt\hbox{$\mathchar"218$}\hss}
     \raise 2.0pt\hbox{$\mathchar"13E$}}}
\newcommand{\lta}{\mathrel{\hbox to 0pt{\lower 3pt\hbox{$\mathchar"218$}\hss}
     \raise 2.0pt\hbox{$\mathchar"13C$}}}

\begin{document}

\titlepage

\begin{flushright}
CERN-TH-210\\
\end{flushright}

{\bf \large Dynamics of Domain Walls for Split and Runaway Potentials}

\rm
\begin{center}
{\bf Z. Lalak$^\dagger$, S. Lola$^{*}$ and P. Magnowski$^\dagger$}
 \end{center}
 \begin{center}
{\small $^\dagger$ Physics Department, University 
of Warsow, Poland \\
$^*$ Department of Physics, 
University of Patras, Greece}
 \end{center}

\begin{abstract}
We demonstrate that the evolution of wall-like inhomogeneities 
in run-away potentials, characteristic of dynamical 
supersymmetry breaking  and moduli stabilisation,
is very similar to the evolution of domain wall networks associated 
with double well potentials. 
Instabilities that would lead to a rapid decay of domain walls
can be  significantly ameliorated by compensation effects between
a non-degeneracy of the vacua and a biased initial distribution, 
which can be  naturally
expected in a wide class or particle physics models that lead
to out-of-equilibrium phase transitions.
Within this framework, 
it is possible to obtain domain walls that live long enough
to be relevant for the cosmic power spectrum and
galaxy clustering, while being compatible with
the observed cosmic microwave background anisotropies.

\end{abstract}


\section{Introduction}

The well-known measurements of the Anisotropies of the Cosmic
Microwave Background Radiation by WMAP \cite{WMAP}, in combination with
the supernovae type Ia observations \cite{SNIa}, 
imply that the evolution of
the universe is dominated by dark energy, and a state parameter that 
is strongly constrained. Among the most popular scenarios 
to explain the data, is to assume the
existence of an inflationary universe with a 
very small cosmological constant
$\Lambda$. In principle, possible contributions to 
dark energy can also be provided from topological defects
which are produced at phase transitions in the universe \cite{vilenkin}.
An interesting possibility, for instance, would have been that,
such contributions are provided by
domain walls  \cite{wallsref} associated with the breaking of 
discrete symmetries (which arise commonly in a wide class of 
particle physics models).

Yet another possibility is that there is no discrete 
symmetry at all. Even then, there could be nearby minima 
separated by a potential barrier, with initial conditions that result
in both minima getting populated with non-zero probabilities. 
In this case we do not have an exact domain wall configuration, 
but (as will become more obvious later) it still makes sense 
to talk about approximate domain walls that interpolate, 
in a broader sense, between 
basins of attraction of nearby local minima. 
And, in fact, we will show that the dynamics and 
evolution of the network of inhomogeneities is very similar in both 
situations - with exact and approximate domain walls. 
As a specific example of the behaviour of the second type we take 
run-away potentials which appear in models of dynamical 
supersymmetry breaking, and  play an important role in 
modern attempts at non-perturbative supersymmetry breaking 
and moduli stabilisation. In fact, it has been 
pointed out by Dine \cite{Dine}, 
that spatially  inhomogeneous field configurations may evolve differently 
in the expanding Robertson-Walker background
than the homeogeneous mode. The inhomogeneities may help to stabilise 
the moduli (such as the dilaton or radion)
at shallow but finite minima, thus avoiding the Steinhardt--Brustein 
\cite{stein-brus} and Buchm\"uller \cite{buchm}
effects. At the same time, 
the energy density inhomogeneities of such
configurations 
may contribute to the 
shape of the power spectrum of CMBR. 
In the case of TeV scale supersymmetry breaking this 
contribution would be unobservable, but the issue of finding the right vacuum 
remains a valid question independently of the mass scale associated with a run-away potential.

\section{Cosmological problems with wall networks and their possible resolution}

There are three main problems in cosmological scenarios that involve a
significant abundabce of domain walls:

(i) Domain walls that could potentially contribute to Dark Energy,
generally predict an equation of state with
$-2/3 < w_X < -1/3$, which would be ruled out from the commonly quoted 
upper bound $w_X < -0.78$ at $95\%$ c.l. 

(ii) Domain walls that could enhance the Cold Dark Matter Spectrum
are in general associated with unacceptably large
fluctuations of the CMBR (Cosmic Microwave Background Radiation), 
for the range of parameters that would have been relevant for 
the formation of structure.
For a horizon-size bubble at a redshift $z_a$, with surface energy $\sigma$,
the generated anisotropies are given by
\begin{equation}
\delta T / T \sim G_{N} \sigma R_{H} (z_{a}).
\label{eq:tin}
\end{equation}

(iii) Domain walls in the simplest class of models that evade 
problem (ii), do not stay around suffiently, in order to produce density 
fluctuations that can sufficiently grow to the observed structures \cite{PRS,Sarkar,Coulson}.

The first problem has in fact been addressed in a very convincing way 
in \cite{ADWRO}, 
where the assumptions made in the
choice of priors of the data analysis have been questioned. 
In fact, it has been shown that,
for  lower values of
the Hubble parameter ($h<0.65$, as indicated by 
Sunyaev-Zeldovich and time delays for gravitational lensing
observations), and for higher values of the matter density
($\Omega_m > 0.35$, in agreement with measurements of the
temperature-luminosity relation of distant clusters observed with
the XMM-Newton satellite), 
domain walls in an inflationary universe
can provide a good fit to the WMAP data.

In previous papers \cite{LR}, \cite{LLOR}, we have 
proposed and tested two main frameworks that may naturally 
arise in standard model extensions for which domain walls can 
lead to the formation of structure, enhancing the standard 
cold dark matter spectrum in an inflationary universe,
while still be compatible with CMBR. These are the following:

a) Schemes where the walls are unstable, due to a non-degeneracy
of the minima of the potential
(as appears naturally in a wide class of superstring models \cite{LR}).
For a large range of possible parameters, the walls  are 
expected to annihilate before recombination.
In this way, although structure can be generated
and subsequently grow
in consistency with the observations, no unacceptable
distortions to the cosmic microwave background radiation
are produced.

b) A second possibility is that,
if one of the minima
of the potential of the scalar field is favoured,
then a biased phase transition occurs.
As a first step, we showed why
such a  bias may be expected in post-inflationary,
out-of-equilibrium phase transitions \cite{LLOR}.
The idea is that, if the interactions of a field are {\em very weak},
this will not be confined at the top of its potential, but
will in fact be centered around a classical value
that will be closer to one of the minima of the potential.
Quantum fluctuations will move it, but, nevertheless, the bias
(offset) will remain. Then, percolation theory indicates that
there is a range of natural initial conditions
for which walls of finite size (and not of horizon size)
are produced inside a sea of the dominant vacuum.
While not very accurate, percolation theory allowed
us to formulate a qualitative picture
of the spatial
distribution of the wall-driven overdensities
in a post-inflationary
universe, and to account for
the whole range of large scale structure observations.
In addition, by studying wall-driven fluctuations at small scales, 
it has been possible to reproduce the observed 
distribution of quasars \cite{LHei}).

Subsequently, elaborate numerical simulations seemed to indicate that
despite the biasing of the minima, the walls either disappear too fast,
or stay around for too long \cite{Coulson}, implying that they have to be very
soft if they are not to lead to unacceptable distortions of the microwave background
radiation. 
In this work, we will give specific examples where 
this need not be the case, firstly in biased double well 
potentials with
non-degenerate minima and secondly, for the runaway potentials that can be
expected in a wide class of supersymmetry breaking models, based on
gaugino condensates.
This complements the literature on the subject and
raises additional possibilities to those that 
have been considered in the recent years
(\cite{Mats} - \cite{Eto}).

\section{Basic Framework for Out-of-Equilibrium, Biased Phase Transitions}

An elaborate numerical study of the dynamics of domain wall 
networks in the case of a scalar field whose potential
has two degenerate minima that occur with the same probability,
has been provided by Press, Ryden and
Spergel, \cite{PRS}, who 
showed that such networks would rapidly evolve into long domain walls
stretching across the universe whose surface area, and, hence, energy density,
persisted for a long time. This resulted 
to a rapid domination of the energy density
of the universe by these walls and to unacceptably 
large distortion in the CMBR. Such an
initial distribution on a lattice can be described statistically using
percolation theory.
On a three dimensional square lattice, there is a critical
probability, $p_{c}=0.311$, above which the associated vacuum will percolate
across the entire lattice \cite{ovr}.
It is easy to see that, by initializing both vacua
with a probability $p=1/2$, both vacua  propagate across the lattice.
Since domain walls lie on the interface between the two different vacua, this
implies the formation of domain walls which extend across the entire
universe.  This gives a clear mathematical  explanation for the Press, Ryden and
Spergel result. However, if for one vacuum $p<p_{c}$, then this
vacuum would form finite clusters in the percolating sea of the other
one. The domain walls would then be small, finite bags 
which would disappear relatively rapidly.
Similar effects would hold in the case of potentials with non-degenerate minima
\cite{LR}.
Here, the true minimum will be at some 
stage energetically favoured, and domain walls will dissappear.

If a phase transition is triggered by fluctuations in a system in 
thermal equilibrium, and the vacua are trully degenerate,
one expects the population probabilities 
of each vacuum to be equal. 
However, non-equilibrium phase transitions, which can occur in
realistic models of the early universe, generically lead to a biased
choice of vacuum state. 
Indeed, an out-of-equilibrium scalar field $\phi$ living on an inflating
de Sitter space, observed over a physical volume $\ell^3$,
breaks into a classical and a quantum piece
\begin{equation}
\phi = \phi_c + \phi_q
\label{1}
\end{equation}
where $\phi_c$ satisfies the classical equation of motion
and $\phi_q$ represents de Sitter space quantum
fluctuations. 
This is illustrated in Figure 1.

\begin{figure}[!h]
\begin{center}
\includegraphics*[height=3cm]{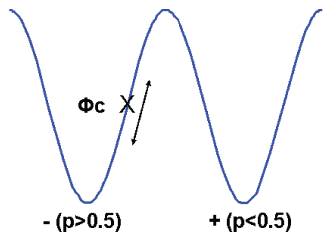} 
\end{center}
\caption{\it Shematic illustration of biased, 
out-of-equilibrium phase transitions.
The initial mean value of the field is shifted towards 
one of the minima, which occurs with a higher probability.}
\end{figure} 

During inflation, and long afterwards, the Hubble term
is very large
compared to the
curvature of the potential and, thus,
to a very good approximation,
$\phi_c = \vartheta$ where $\vartheta$  is
an arbitrary
constant (to
next order, there is a tiny damped  velocity
$\dot \phi_c
\sim \frac{V}{H \upsilon}$).
On the other hand, $\phi_q$
represents
quantum
fluctuations  of the scalar field in de Sitter space.
These fluctuations
result in the formation of a weakly inhomogeneous
quasi-classical  random field. After inflation ends, the FRW horizon,
$\ell_c=1/H$, grows and fluctuations with scales less than the
horizon are smoothed out. Thus $\ell_c$ acts as an  UV cut-off in
the momentum distribution of this random field. In a spatial
region of length $l$,  the distribution
of the fluctuations around $\vartheta$ can be calculated and is
given by \begin{equation}
P(\phi)=\frac{1}{\sqrt{2 \pi} \sigma_{\ell}} \exp{(-\frac{(\phi -
\vartheta)^2}
{2 \sigma^2_{\ell}})}
\label{prob}
\end{equation}
where
\begin{equation}
\sigma^2_{\ell} = \frac{H_i^2}{4\pi^2} \ln \left(
\frac{\ell}{\ell_c} \right)
\label{3}
\end{equation}
(and, as has been discussed in \cite{LLOR}, for fields produced towards the
end of inflation, one can ensure that
 the longwavelength
components in the
Fourier
decomposition of the random field $\phi_q$ do not introduce 
unacceptable for our discussion correlations
between the values of the random field at distant points).

Clearly, such transitions can only occur in a system that is too
weakly coupled to achieve thermal equilibrium. 
A number of such biased
transitions have been investigated, includingy those occurring in very
light scalar fields in deSitter space \cite{ovr}.

The two disconnected minima are
denoted by $(+)$ and $(-)$ respectively.
At the phase transition, the field has some finite correlation length 
(like the inverse Ginzburg temperature in 
case of a transition triggered by thermal fluctuations) over which 
the post-transition vacuum
is chosen coherently, denoted by $\Lambda$.
One can approximate the initial 
spatial structure of the vacuum produced during the transition
by first dividing space into cells of volume $\Lambda^{d}$, where $d$ is the
dimension,
and second, by assuming that choices of the new vacua are made independently 
in each cell, giving the $(+)$-vacuum with probability $p$ and 
the $(-)$-vacuum with probability $1-p$,where $0\leq p\leq 1/2$. 
Whenever the vacua in neighboring cells are different, 
a domain wall will form 
which interpolates between them, and so, typically,
a complicated spatial network of domain
walls will form.
 
Of course, an arbitrary spatial superposition of domain walls,
such as  that produced by the mechanism described above, 
is not a solution of the equations of motion and
cannot be stable. However, such a superposition
represents physical initial conditions, the
subsequent evolution of which is governed  by the dynamics of the
theory. Subject to this dynamics, the initially static domain
walls  acquire non-zero velocities, oscillate under their
surface tension, and interact with one another. This will be discussed in subsequent sections,
where we will summarise the main ingredients, 
but also motivation
and naturalness of out-of-equilibrium, 
biased phase transitions, 
where the bias can come from: \\
 (i) small differences in 
the energy of the minima of the potential, and \\
(ii) different probabilities to reach these minima.

\section{Dynamics of the Scalar Field and Wall Network}

How do we study the behaviour of a  ``biased network''?
The scalar field is initialized by randomly setting
it equal to $-\phi_0$ or $+\phi_{0}$ at
each lattice site, with bias probability $p$ for $+\phi_0$ 
and $1-p$ for $-\phi_0$ with $0\leq p \leq 1/2$.  The
lattice resolution corresponds to the initial field correlation length,
and on physical scales above the resolution cut-off the field will 
have a white noise power spectrum (yielding results similar to percolation theory). 

In this section, we follow the study presented in \cite{PRS} and subsequently extended in
\cite{Coulson}. The dynamics of the scalar field, $\phi$, is
determined by the equation of motion. This has the form

\begin{equation}\label{r_ruchu_t}
\frac{\partial^{2}\phi}{\partial t^{2}} + \frac{3}{a} \frac{\partial
a}{\partial t} \frac{\partial \phi}{\partial t} -
\frac{1}{a^2}\nabla^{2}\phi = - \frac{\partial V}{\partial \phi}.
\end{equation}

and, introducing the conformal time $\eta$
(with  $d\eta = \frac{dt}{a(t)}$), it becomes

\begin{equation}
 \frac{\partial^{2}\phi}{\partial \eta^{2}} + 2\frac{\partial a}{\partial t}
 \frac{\partial \phi}{\partial \eta} - \nabla^{2}\phi =
 -a^{2}\frac{\partial V}{\partial \phi}.
\label{eqmot}
\end{equation}

In the above,
$a$ is the scale factor of the universe ($a \sim \eta$
in the radiation era, and $a\sim \eta^2$ in the matter era),
$V$ is the scalar potential  and the spatial gradients are with 
respect to co-moving co-ordinates.   
Then:
\begin{equation}
    \rho = \frac{1}{2}\frac{\partial^{2}\phi}{\partial t^{2}}
    + \frac{1}{2a^2}|\nabla\phi|^2 + V(\phi),
\end{equation}
\begin{equation}
    p = \frac{1}{2}\frac{\partial^{2}\phi}{\partial t^{2}}
    - \frac{1}{6a^2}|\nabla\phi|^2 - V(\phi).
\end{equation}

The scalar potential
determines the topology of the vacuum manifold.  
A typical choice is a
$\phi^4$ potential
\begin{equation}
\label{eq:potential}
V(\phi)= V_0 \left(\frac{\phi^2}{\phi_0^2} -1\right)^2
\end{equation}
with the two degenerate vacua, $\phi=\pm\phi_0$, separated by a
potential barrier $V_0$. 

One can define a physical domain wall thickness $w_0$ given by
\begin{equation}
\label{eq:thick}
w_0 \equiv \pi \frac{\phi_0}{\sqrt{2V_0}}.
\end{equation}
The ratio of the wall thickness to the horizon size (${\cal H}^{-1} =
\left( \frac{1}{a} \frac{\partial a}{\partial \eta} \right)^{-1}$)
at the time of the phase transition 
\begin{equation}
\label{eq:eta0}
W_0 \equiv \frac{w_0}{a(\eta_0)} \frac{1}{\eta_0}
\left. \frac{ d \ln{a}}{ d \ln{\eta}}\right|_{\eta_0}
\end{equation}
then sets $\eta_0$, the conformal time of the phase transition
and the time at which we begin the simulation (one needs
the walls to be thinner than the horizon in order to study their
dynamics, namely ~$w_0 \ll H^{-1}$).

Here we assume that the expansion is dominated by some smooth component,
filling the universe. The equation of state of this component is
 $p = \alpha \rho$. This gives
$ a(t) = a_0 t^{\frac{2}{3(\alpha+1)}}$, and 
$d\eta = \frac{dt}{a(t)}$,  $\eta =
t^\frac{3\alpha+1}{3(\alpha+1)}$. Also,
$$
    a(\eta) \sim \eta^\frac{2}{3\alpha +1} \sim \eta ^\omega.
$$

The equation of motion for static domain walls is
\cite{PRS}
\begin{equation}
    \phi = \phi_{0}\tanh \left[\frac{\sqrt{2V_0}}{\phi_0}a(z-z_0) \right]
     \equiv \phi_0 \tanh \left[\frac{a(z-z_0)}{w_0}\right]
\end{equation}
and for non-static (boosted with a velocity $v_0$):
\begin{equation}
    \phi = \phi_{0}\tanh\left[\frac{\gamma_0
    }{w_0}a(z-z_0-v_{0}t)\right],
\end{equation}
where $\gamma_0 \equiv (1-a^2 v^2_0)^{-\frac{1}{2}}$.

The energy and surface density of the walls is
\begin{equation}
    \rho(z) = \frac{\gamma^2_0 V_0}{2}
    {\rm{sech}}^4\left[\frac{\gamma_0}{w_0}a(z-z_0-v_0t)\right]
\end{equation}
\begin{equation}
    \sigma = a\int_{-\infty}^{+\infty} \rho(z)dz = \frac{2\gamma_0 V_0
    w_0}{3}.
\end{equation}

Finally, during the expansion, the velocity of the wall changes 
according to
\begin{equation}\label{spowalnianie_scian}
    \gamma(t)v(t) \sim a(t)^{-4}.
\end{equation}

Let us now pass to run-away potentials, of the form
$$ V(s) = \frac{1}{2s} {\left( A(2s+N_1) e^{-\frac{s}{N_1}} - B(2s+N_2) e^{-\frac{s}{N_2}}\right)}^2.$$
In this case we do not have an analytic solution of the domain-wall type, 
interpolating between finite and run-away minima. However, 
we may still use the domain-wall language to describe the 
distribution of energy and topology of the vacuum. 
We shall call as `domain walls' the non-equilibrium configurations which appear 
on the lattice as joining field values in naighbouring lattice sites; then,
we can identify the position of these generalised walls as the link between 
the lattice sites occupied by different vacua. In the case of the run-away vacuum,  
we shall simply determine whether a field value at a given site 
belongs to the classical domain of attraction of that vacuum. 

To further develop an intuitive feeling about the evolotion 
of the system we shall define a domain wall width, demanding that 
it should correspond to a distance in configuration space 
over which the field gradient is of the order of the 
potential energy of the local 
maximum that separates the vacua.  In other words, 
$$\Delta = \frac{|\phi_{max} - \phi_{min}|}{\sqrt{V(\phi_{max}) - V(\phi_{min})}},   $$
where $\phi_{max}$ denotes the position 
of the maximum separating the domains of 
attraction of the finite minimum, $\phi_{min}$, and 
of the run-away minimum ($\phi \rightarrow 
\infty$). 

\section{Numerical procedure: description and testing}

The numerical implementation of the equation of motion 
(\ref{eqmot}) is non-trivial. 
It involves  discretisation of the equation of motion (Appendix I), which allows
treating the domain wall network numerically. Moreover, the calculations are very 
time-consuming, unless an optimisation of the time step is applied. We propose 
in Appendix II such a technique, which very significantly improves the
efficiency of the code, and allows us to go to larger lattices and 
higher accuracies. The role of the size of the lattice is discussed in Appendix III.

There are  additional considerations to be made:
to start with, the factor $a^2$ on the right hand side of the equation
of motion
makes the effective potential barrier grow with the expansion.
The result is that, in comoving coordinates, the width of the walls
decreases like $a^{-1}$, which is  $\eta^{-1}$ 
for radiation dominated and 
$\eta^{-2}$ for a matter dominated universe.
This implies that, on any reasonably sized
grid, it is impossible to ensure that the walls would be visible on
the lattice to the end of a calculation, when the horizon size
is roughly the grid size.  
To appropriately represent walls, their width  should be
of the order of a few lattice sites during the whole simulation
(in our case, we require the walls to be about five lattice 
sites wide since, if they become too wide, we lose the resolution).

However, we know that the dynamics of the walls
does not depend on their width once they get created and separated from each other
\cite{PRS}, while the total surface energy and surface tension 
also do not depend on the width.
As a result,  one can consider a
generalization of the equation of motion, which may force the walls
to maintain a constant co-moving thickness while otherwise
not altering their dynamics.  This modified equation is
\begin{equation}\label{r_ruchu_uog}
 \frac{\partial^{2}\phi}{\partial \eta^{2}}
 + \frac{\alpha}{\eta} \left(\frac{d\ln{a}}{d \ln {\eta}} \right)
 \frac{\partial \phi}{\partial \eta}
 - \nabla^{2}\phi = -a^{\beta}\frac{\partial V}{\partial \phi},
\end{equation}
and, for $\alpha=\beta=2$, we recover the initial equation of motion.

If $\beta=0$ the walls will have constant comoving width.
This choice does alter the scaling of the adiabatic effects of the Hubble
expansion ($\beta =0$), but this effect can be compensated by a proper choice of 
$\alpha$. It turns out that \cite{PRS}
\begin{equation}\label{skalowanie_fi}
    \langle \phi - \phi_0 \rangle_{\rm{rms}} \sim a^{-\frac{\alpha}{2}
    - \frac{\beta}{4}}.
\end{equation}
Thus, for $\beta = 0$, we have to set $\alpha = 3$
to have the same scaling of the deviation of $\Phi$ and $\Phi_{0}$.
In addition,
\begin{equation}\label{spowalnianie_uog}
    \gamma v \sim a^{-\alpha -\frac{\beta}{2}}.
\end{equation}

Having obtained a consistent set of equations, the next step is to understand
the relevant range of initial conditions and parameters.
As in previous works,  we will set 
$\phi_0=1$.  The scalar field initial conditions 
are then chosen using the prescription described above for various
bias probabilities,
$p$. That is, in the following we will use percolation theory with allowed
field
values of $\pm1$ at any lattice site. It is of interest to compare the 
evolution of 
the network initialized with two-point initial conditions with
those initialized with various continuous distributions. We have done 
this for a 
uniform distribution which gives probability $1-p$ of choosing $\phi$ between 
$-1$ and $0$ and probability $p$ of choosing between $0$ and $1$, 
and with a gaussian
distribution $P(\phi)$ such that $\int_{0}^{+\infty} d\phi P(\phi)=p$. 
In general, the surface area of the initial network, measured at $\eta=\eta_0$,
is largest in the case of the two-point, percolative distribution. 
However, after a few steps of dynamical 
evolution the network stabilizes, and its important characteristics, 
such as surface energy or kinetic energy and their time evolution, 
become indistinguishable for a fixed bias $p$. Hence, the sharp initial 
conditions
of percolation theory also give a good approximation to initial conditions
softened by smooth distribution functions. This justifies the use of the pure
two-point percolation theory initial conditions in this paper.

We will set the initial field ``velocity'', $\dot{\phi}$, to be zero 
everywhere on the lattice (in previous work 
the results were found to be insensitive 
to small initial velocities with respect to the energy of the barrier
\cite {Coulson}; this
was done by repeated simulations with $\dot{\phi}$
chosen from a uniform distribution of velocities between $-1$ and $+1$
($\sim {\cal O}\left(\phi_0/\eta_0\right)$).

Simulations are run in the radiation dominated epo\-ch, with
an initial time, $\eta_0=1$, unless stated otherwise.
We chose a wall thickness $w_0=5$, and a ratio $W_0=5$ {(see (11))}.
This value is used to ensure that the wall thickness is well
above the lattice resolution scale (recall $\Delta x=1$), while also
ensuring that for most of the dynamical range of the simulation, the wall--wall
separation exceeds the wall thickness.  

We tested our code with the simplest case
one can study, namely the double well potential, for which 
\[
V(\phi) = V_0 \left( \left(\frac{\phi}{\phi_0}\right)^2 -1\right)^2 
\]
($ V(\phi) = m \phi^2 +
\lambda \phi^4$,  $m = -2V_0 {\phi_0}^2 $, 
$\lambda = V_0$).

Looking at the potential and the first derivative (Figure 2) we see that, while the
potential is symmetric (even), the first derivative is asymmetric (odd):

\begin{figure}[!h] 
  \begin{center}
  \epsfbox{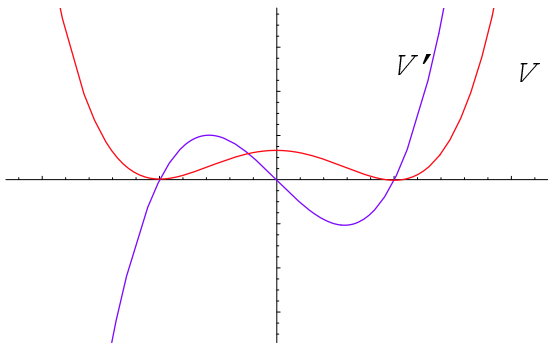}
  \end{center}
  \caption{\it Shape and first derivative of the double well potential}
  \label{etykieta4}
\end{figure}
This is a special situation, and this symmetry is violated in the case of the 
exponential potential discussed later on. 
  
For the numerical part, we followed 
\cite{PRS} and \cite{Coulson}, setting the values of the
parameters $\alpha$ and $\beta$ to reproduce the original equation
of motion, equation~\ref{eqmot}.  In the runs the domain
walls maintained a constant physical rather than co-moving
thickness, and so problems of available dynamic range become
important.  The prescription used is as follows:
The simulation is run on a $512\times 512$ lattice, with our
standard field initial conditions and $p=0.5$.  Using a wall thickness
of $w_0=25$, the initial conditions were evolved 
with the standard parameter values of $\alpha=3$ and $\beta=0$,
 until a time when
the wall-wall separation exceeded the wall thickness; that is, a
time $\eta=2w_0$.  The equation of motion was then changed, to set
$\alpha=\beta=2$ until a time $\eta=250$.  The co-moving thickness
of the walls at the end of the simulation was then $5$ lattice sites.
To compare, the run was repeated, this time leaving the standard
parameter values fixed throughout the simulation.  

Looking at the  evolution of the energy density
of the network of domain walls in three dimensions,
(in a   radiation dominated epoch), we confirm the following results
\cite{Coulson}.

$\bullet$
For  $p<0.311$, the critical threshold of percolation theory,
 only one vacuum percolates the 
lattice, and isolated bags of one vacuum 
are to be found in a percolating sea of the
more dominant vacuum. These bags rapidly decay under their
surface tension.

$\bullet$
For $0.5>p>0.311$ both vacua percolate, leading to
an initial network of infinite
(lattice sized) domain walls.  However, 
these also rapidly decompose into vacuum
bags which then decay.  

$\bullet$
Only in the exact $p=0.5$ case is long-term scaling
seen, dominating the energy density
of the universe only in the case of $p=0.5$.

What is thus seen, is that, 
a
network of domain walls forming  well before matter-radiation
equality, can be 
sufficiently massive to contribute significantly 
to large scale structure formation on comoving scales
less than $\sim 20\ Mpc$. However, such a network
will  decay before photon decoupling.

An immediate question is whether superhorizon fluctuations
can be of any relevance. In principle,
such a possibility can arise, particularly in a universe with 
a significant hot dark matter component, and
scale invariant primordial perturbations
induced by an earlier inflationary epoch. However, 
similarly to \cite{Coulson}, we found that,
for much of the range of biases, 
wall networks turned out to be cosmologically innocuous,
as their energy density
exponentially decays with a characteristic time of only
a few expansion times.  Simulations were made in both 
3-dim and 2-dim; in the
later case walls stay around longer, but still,
for any significant scaling
of the network before the ultimate exponential decay,
finetuning of $p$ close to $1/2$ is required.

In principle, one has to consider also the possibility
of having  bias in the initial velocities.
However, for the double well potentials, modifications from such effects are 
negligible. The reason is that in this case the field is perfectly reflected
to the minima by the external barriers; this however is not what happens for
the runaway potentials that we will proceed to discuss, as well as for periodic
potentials. 

\section{Biased Potentials with non-degenerate minima}

In the previous section, we summarised the expectations for 
potentials with degenerate minima. However,
the behaviour of domain walls can change radically, in the case that the
minima of the potential are unstable.
Before studying the domain wall dynamics that are to be expected,
we would first like to discuss the naturalness of such solutions.

This problem has been studied extensively in \cite{LR}, 
where specific realisations of such scenarios have been proposed:
In realistic standard model extensions, and particularly
in superstrings
there are usually several discrete groups
$Z_N$. The fields 
in the theory then transform as $\alpha^r$, r=0,1,..,N-1, where
$\alpha$ is the $N^{th}$ root of unity, and the effective
potential is constructed from $Z_N$ invariant combinations of the
fields. In non-supersymmetric models, for example,
the Lagrangian of a complex scalar field $\Phi$, transforming as
$\alpha$ has the form:
\begin{equation}
L = \partial_{\mu} \Phi \, \partial^{\mu}
\Phi^{*} 
+ \mu^{2} \mid\Phi\mid^{2}
- \frac {\lambda \mid\Phi\mid^{4}}{4}
+\lambda^{\prime}(\frac{\Phi^{N}}{M^{N-4}} 
+ \frac{\Phi^{* N}}{M^{N-4}})+..
\label{eq:e1b}
\end{equation}
where the  coupling $\lambda$ is made real by absorbing its phase
in the field $\Phi$. 
The coupling $\lambda'$ is of order unity.
The non-renormalisable terms of dimension
$>4$ arise because we have an effective field theory generated
by physics at some (high) scale, M.  

If $\mu^2$ is positive,
the effective potential for $\Phi$ has a
minimum for non-vanishing value of the modulus and leads to
spontaneous symmetry breaking of the discrete $Z_N$ group. In
this case it is convenient
to reparametrize $\Phi$ as
\begin{equation}
\Phi = (\rho + \upsilon) 
\,e\,^{i \,\theta / \upsilon+\alpha },
\label{eq:e2}
\end{equation}
where $\upsilon e^{i \alpha}$ is the v.e.v.  (vacuum expectation
value) of $\Phi$, while $\rho$ and $\theta$ are real scalar
fields. The potential of the field $\theta$ is then \cite{LR}
\begin{equation}
 V ( \theta ) =  \frac{2 \lambda^{\prime }\upsilon^{N}}{M^{N-4}}
\,
\cos(\frac{\theta N}{\upsilon } + N \alpha)
\label{eq:e3}
\end{equation}
and, the value of the pseudo-Goldstone mass is given by
\begin{equation}
m^{2} = \frac{N^{2} V_{0}}{\upsilon^{2}}
\cos(N \alpha)\, \, ,
\,\,\,\,\,V_{0} \equiv \lambda'
\frac{2 \upsilon^{N}}{M^{N-4}}.
\label{eq:e4}
\end{equation}
The potential of equation~(\ref{eq:e3}) has an N-fold degeneracy
corresponding to $\theta /v\rightarrow\theta /v+2\pi /N$
\footnote{Pseudogoldstone bosons, due to their very light mass and negligible interactions,
may very naturally give rise to late phase transitions
\cite{late}.}.

How can this degeneracy be lifted?
So far we have discussed 
domain walls which are expected to arise 
from the potential of a {\em single} scalar field $\Phi$.
However additional scalar fields are also present.
Then, if, as is likely, the interactions between the fields
cause
more than one field transforming non-trivially under the
discrete symmetry group to acquire a vev,
it is possible to generate a situation in which the vacuum 
degeneracy is apparently lifted:
If one of these fields acquires its vev
before or during inflation the observable universe will have a
unique value for its vev. After inflation the effective potential
describing the remaining fields may still have an {\it
approximate} discrete symmetry but the vacua will not be exactly
degenerate. 

To illustrate this, for instance in non-sypersymmetric models,
 consider adding to the above theories a second
field $\Phi^{\prime}$. If $\Phi$ transforms as $\alpha^{m}$
and $\Phi'$ as $\alpha^{n}$ under the symmetry
group and assuming that $n \geq m$ and 
$N/n$ is integer 
in order to simplify the analysis,
the potential is \cite{LR}
\begin{equation}
V(\theta) =\sum_{r=0}^{N/n} \left[2 
\frac{\upsilon^{(N-nr)/m}
\upsilon^{\prime r}}{M^{(N-(n-m)r)/m-4}} 
\cos \left(\frac{r
\theta^{\prime}}{\upsilon^{\prime}}+
\frac{(N-nr)}{m}\alpha+r\beta\right)
\right].
\label{eq:asy2}
\end{equation}

Clearly there will be a dominant term, however the subdominant ones
will {\em slightly} split the degeneracy of the potential.

Having summarised the model building aspects, the next question
is whether the domain walls to be expected in these theories can
be of any relevance for structure formation. 

In the case of non-degenerate minima, we expect that there
is a critical scale at which the 
loss in surface energy from collapsing horizon-size bubbles 
becomes similar to the gain in volume energy 
by passing to the true minimim.
At some stage, the true minimum will be favoured at all horizons,
and walls will disappear.
In this case, for subhorizon fluctuations,
we will have a maximum scale, corresponding to the
size of the horizon $R$ at the time that the walls disappear 
(which today would correspond to
$R^{\prime}=R_{H_0}/\sqrt{1+z_{a}} $). 
These could in principle generate structure for 
the smaller angular scales of WMAP.
The question of course is what is the situation 
regarding superhorizon fluctuations, and whether 
these can give any structure at the largest scales
(COBE, and the largest scales of WMAP).

The time when walls dissappear is specificed by a redshift 
$z_a$, which is a calculable quantity, even before passing to
any numerical simulation.
 For non-degenerate vacua there is always
a critical bubble radius above which it is energetically
favourable for the bubble of true vacua to expand gaining more
volume energy than is lost in surface energy. Once the horizon
exceeds this critical radius bubbles of the true vacuum will
expand everywhere at the speed of light to fill the whole
universe and this occurs at the same time in all horizon
volumes \cite{coleman}. 
Then, if the non-degeneracy of the potential
is measured by a factor $\delta \rho \approx 
\frac{\sigma}{R}$.
For instance, if walls decay during matter dominance, this
 determines the redshift $z_a$ to be 
\begin{equation}
z_{a} = \left(\frac{\delta \rho R_{H_{o}}}
{\sigma}\right)^{2/3} - 1 \, .
\label{eq:zz}
\end{equation}

In what follows, we combine biasing with non-degeneracy of minima,
a situation that, according to the above, can arise for effective
potentials generated by several weakly interacting scalar fields.
Then, if the  minimum with the highest probability
has a higher energy than the second one, we have two
competing effects, which  can allow modifications from 
previous results in the literature.
This is illustrated in Figure 3.

\begin{figure}
\begin{center}
\includegraphics*[height=3cm]{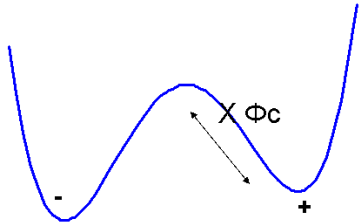} 
\end{center}
\caption{\it Shematic illustration of biased, out-of-equilibrium 
phase transition, in a potential with non-degenerate minima.
The field is shifted towards the false vacuum.}
\end{figure} 

To undestand these effects, we perform a numerical similation.
A simple and generic parametrisation of the non-degeneracy of the potential
is obtained by  adding to $V(\phi)$ a term linear
in $\phi$, namely

$$V(\phi) \rightarrow V(\phi) - \epsilon \, V_0 \, \phi \; .$$

The monitoring of the extrema of the potential is shown
in Appendix IV.  During the evolution, 
the field ``feels'' only the derivative
of the potential and the gradient of the field, thus 
it has the tendancy to evolve towards the place
where the magnitude of the derivative is larger. This may in principle compensate
the biasing of the initial distribution; in fact,
 the two effects can be combined in such a way, that one produces a
quasi-stable network, which is shown in Figure 4.

\begin{figure}
\hspace*{2.0cm} 
\vspace*{0.5 cm}
\includegraphics*[height=6cm]{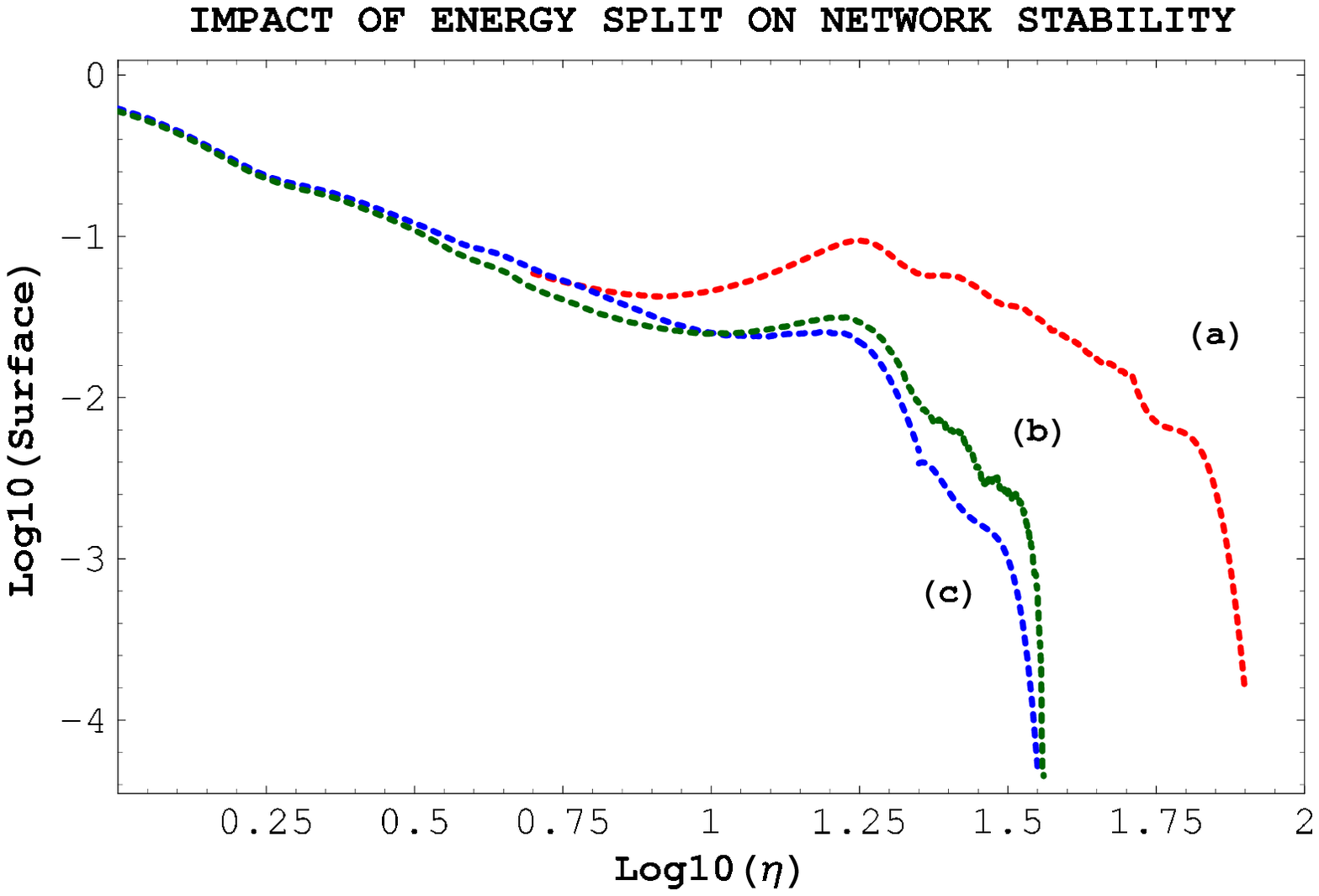} \\
\hspace*{2.0cm} 
\vspace*{0.5 cm}
\includegraphics*[height=6cm]{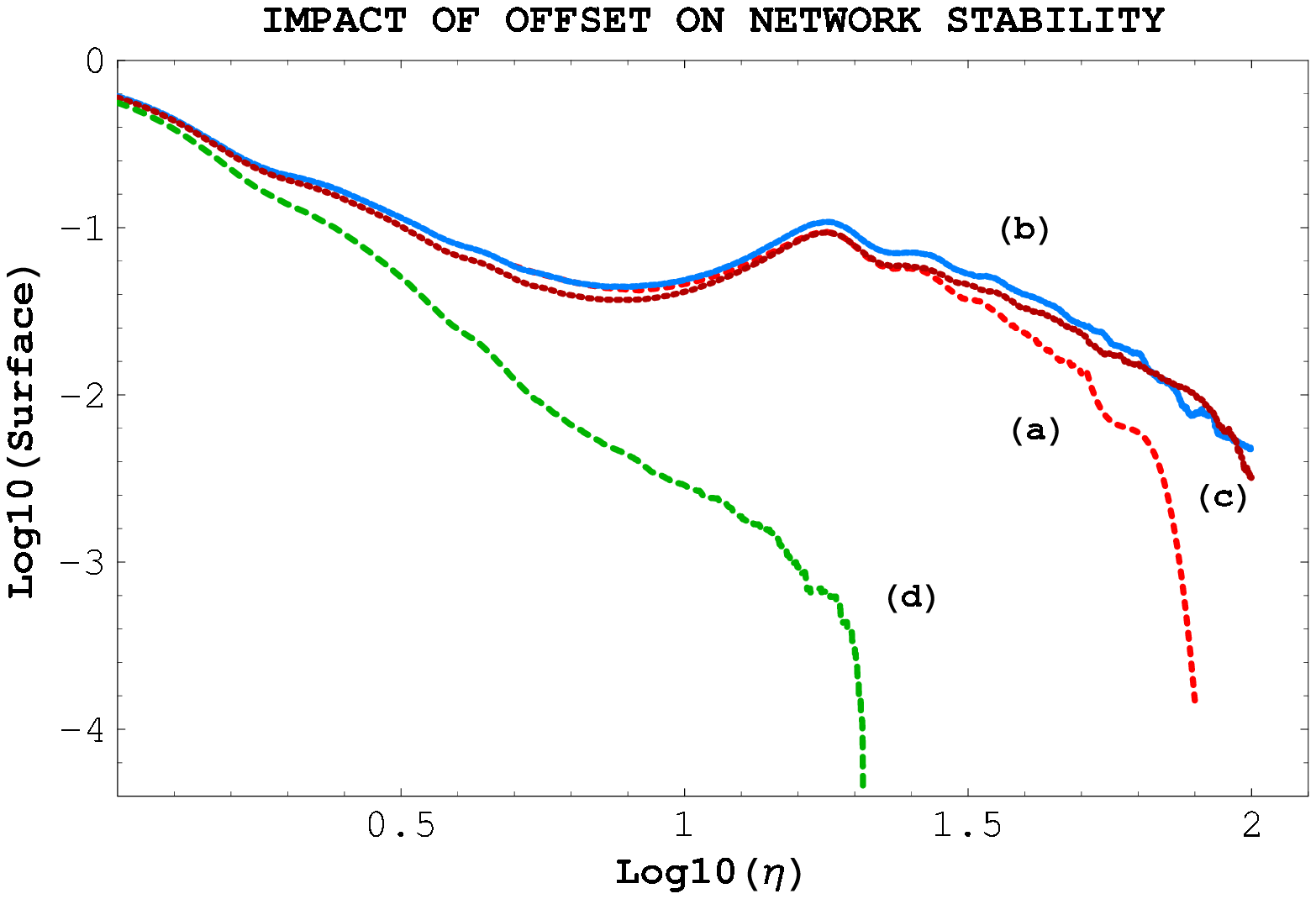} \\
\hspace*{2.0cm} 
\vspace*{0.3 cm}
\includegraphics*[height=6cm]{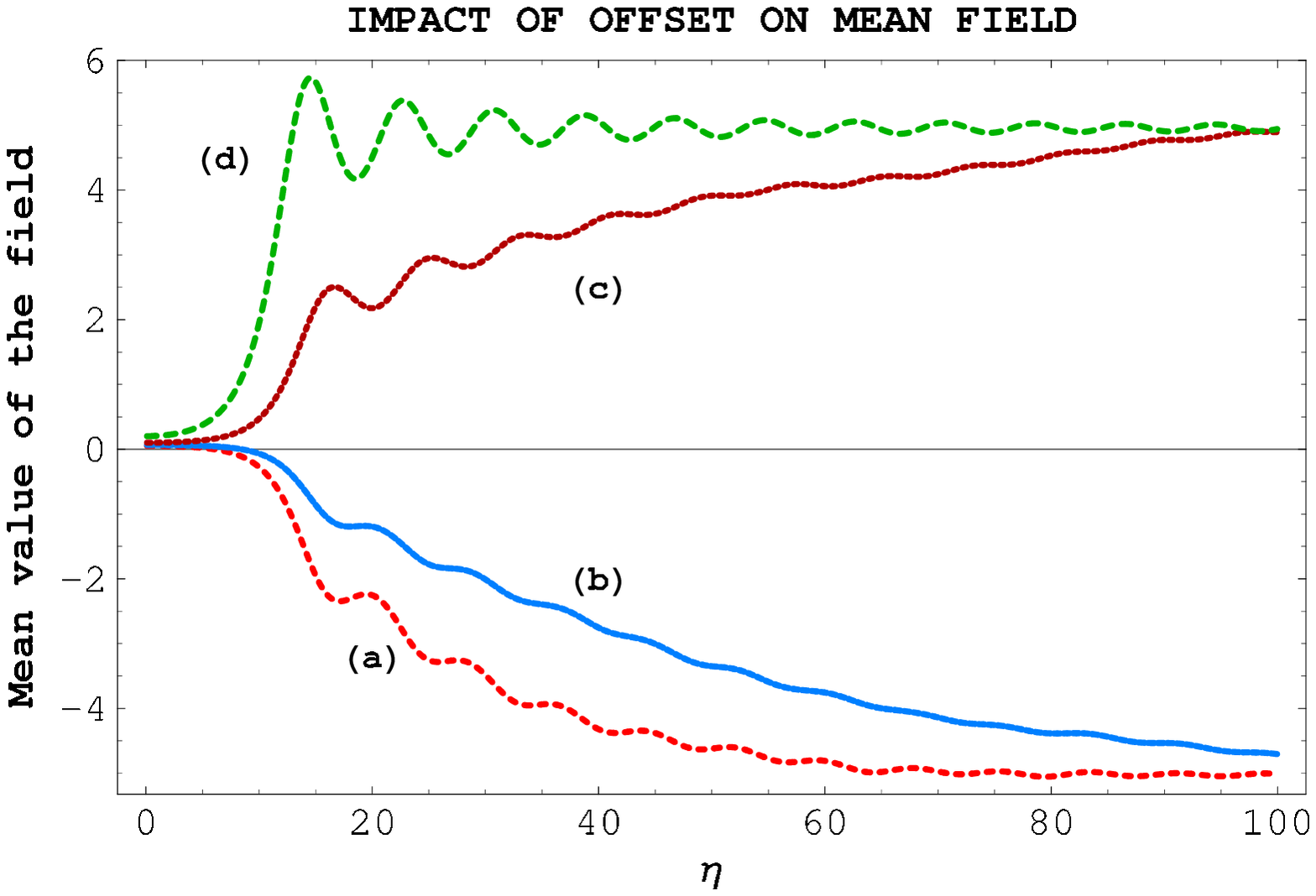}  
\caption{\it Domain wall evolution in potentials with 
non-degenerate minima and a 
bias in the initial mean field distribution.}
\label{pro:fig} 
\end{figure} 
In the upper panel of Figure 4, we present the 
evolution of the surface energy of
the walls, as a function of the conformal time
$\eta$, for three different cases  with the bias in the initial distribution corresponding to 
$p_{-}=0.47$: \\
(i) The upper curve, (a), corresponds to the case where the effect 
of the non-degeneracy of the minima,
parametrised by $\epsilon=-0.012$,  is partially cancelled by
the bias in the initial field disribution. \\ 
(ii) The middle curve, (b), stands for the case of 
degenerate minima. \\
(iii) The lower curve, (c), denotes the case with a 
higher non-degeneracy of the minima, parametrised by
$\epsilon = -0.02$. As expected, this 
choice leads to a rapid disappearence of the walls.

The middle picture shows the behaviour of the 
surface energy of the network 
for different initial distributions and a fixed value of the
non-degeneracy of the minima ($\epsilon = -0.012$). \\
(i) Curve (a) is the same as above, with a bias given by $p_{-}= 0.47$. \\
(ii) Curve (d) corresponds to a bias $p_{-} = 0.39$, that is  
a probability to occupy the left (lower) vacuum equal to $0.39$.\\
(iii) In curves (b) and (c) the initial bias of the 
distribution is given by $p_{-}=0.46$ and $p_{-}=0.44$ respectively. 
In these cases the offset has been tuned in such a way that it compensates the effect of non-degeneracy. 
One should note that the necessary tuning is of the order of a few percent,    
enhancing the a'priori  probability of occupying the right vaccuum by about 5\%. 

The lower panel shows the evolution of the mean value of the
scalar field during the decay of the wall network, for the same choice of parameters
as in the middle panel. Curves (a), (b), (c), (d) correspond to 
the respective ones in the middle panel.

The outcome is that we were able to realise a quasi-stable network 
by combining two competing effects. 

\section{Runaway potentials}

\subsection{Theoretical Motivation and Description of the Potential}

Runaway potentials arise commonly in theories of dynamical 
supersymmetry  breaking, hence their collective 
dynamical properties deserve a careful study. For the 
purpose of this paper we assume the potential in 
the form  

$$ V(s) = \frac{1}{2s} {\left( A(2s+N_1) e^{-\frac{s}{N_1}} - 
B(2s+N_2) e^{-\frac{s}{N_2}}\right)}^2,$$

\begin{figure}[!h] 
  \begin{center}
  \epsfbox{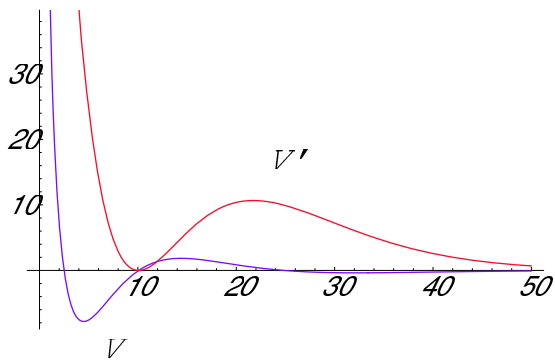}
  \end{center}
  \caption{\it Shape and first derivative for the runaway potential.}
  \label{etykieta4b}
\end{figure}

Its shape and first derivative are plotted in  Figure 
\ref{etykieta4b}.

If we would like to identify the field $s$ with a stringy modulus, say a dilaton,
 then we would have to define $ s = e^{\sqrt{2} \phi} $, since $\phi$ defined in such a way is
canonically normalised. This makes the potential above a doubly-exponential function of $\phi$. 
However, in what follows we shall simply assume that the K\"ahler potential for $s$ is canonical - 
this simplification doesn't introduce qualitatively new features in the simulations.
In general, one can see that the doubly-exponential steepness of the potential makes the evolution more  
sensitive to the changes of the offset and the width of the distribution. 

The degeneracy of this potential can be lifted by adding
a term of the form $\frac{\epsilon}{s^2}$.
The position of the extrema of the potential, and the way they are monitored
in our simulations are discussed in Appendix IV.

We have studied runaway potentials for the parameter set of Table 
\ref{ta1},
which corresponds to a weakly coupled vacuum. The expectation balue of $s$ 
($<s> \sim 10$) corresponds to inverse square of a gauge 
coupling in a supersymmetric Yang-Mills theory. 

\begin{table}[!h]
 \centering
\begin{tabular}{|c|c|c|c|c|c|c|}
  \hline
  A & B & $N_1$ & $N_2$ & min & max & $w_0$ \\
  \hline
  1.0330 & 1.1950 & 10.0 & 9.0 & 10.075 & 21.729 & 3.567 \\
  \hline
\end{tabular}
\caption{\it Runaway potential parameter set used in our simulations.}
\label{ta1}
\end{table}

In Table 1,  min-max denote the minimum and maximum of the potential,
and $w_0$ gives the width of the (approximate) domain wall.
The initial  conditions are controlled by the position of the 
mean value of the initial distribution,
 $\langle \phi \rangle = max + w_0 \gamma$,  
and by the initial width of distribution, $\sigma_\phi = w_0 \gamma'$, 
where $\gamma, \, \gamma'$ are real numbers to be discussed later on. 

\subsection{Numerical simulations for Runaway Potentials}

Runaway potentials are in principle more complicated to study than
the double well ones,
since they have more intrinsic instabilities.
This implies that several of the assumptions made for the simplest
potentials, have to be re-considered. This by itself is an interesting
problem and will allow to understand the level of validity of the
results in potentials that are to be expected in theoretically motivated
models.

\underline{Modification of equation of motion}

The first step is to analyse what is the effect of the modification
of the equations of motion, (by taking
$\alpha =3$ and $\beta=0$), in order to maintain a constant
comoving thickness for the walls, while
maintaining the condition for conservation of the wall momentum.
The condition for momentum conservation is
$\beta = 6 - 2\alpha$. 

\begin{figure}[!h] 
  \begin{center}
\includegraphics*[height=7cm]{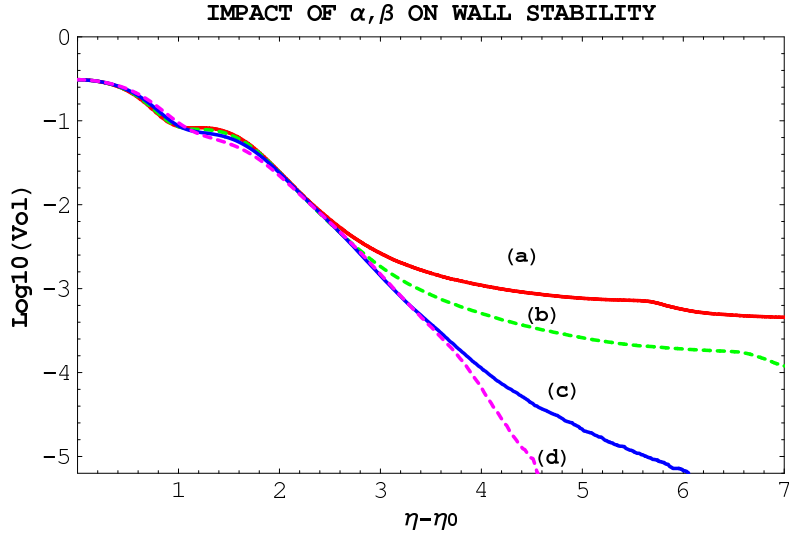} 
  \end{center}
 \centering
\begin{tabular}{|c|c|c|c|c|}
  \hline
  & (a)  & (b)  & (c)  & (d) \\   
 \hline  \hline  
  $\alpha$ & 1.8  & 2.0  & 2.4  & 3.0   \\
  \hline
  $\beta$  & 2.4  & 2.0  & 1.2  & 0.0   \\
  \hline
\end{tabular}
  \caption{\it $\log_{10}(\rm{Vol(L)})$ 
as a function of  $\eta - \eta_{start}$,
for four different combinations of $\alpha$ and $\beta$.}
  \label{albe}
\end{figure}

To do so, we perform simulations for intermediate 
values of $\alpha$ and $\beta$ to see how this
modification changes the evolution of the wall network.
The results are  shown in Figure \ref{albe}.

In general, the larger $\beta$ is, 
the slower is the rate of disappearence of the
false vacuum. When $\beta>1.2$, the bubbles of the vacua become stable.
This is because the effective potential barrier grows with time with respect to the
gradients, so, they cannot overcome the potential. When 
$\beta < 1.2$, the presure of the dominant vacuum takes over.

\underline{Study of the equation of state}

\begin{figure}[!h]
\hspace*{-1.5cm}
\includegraphics*[height=5.5cm]{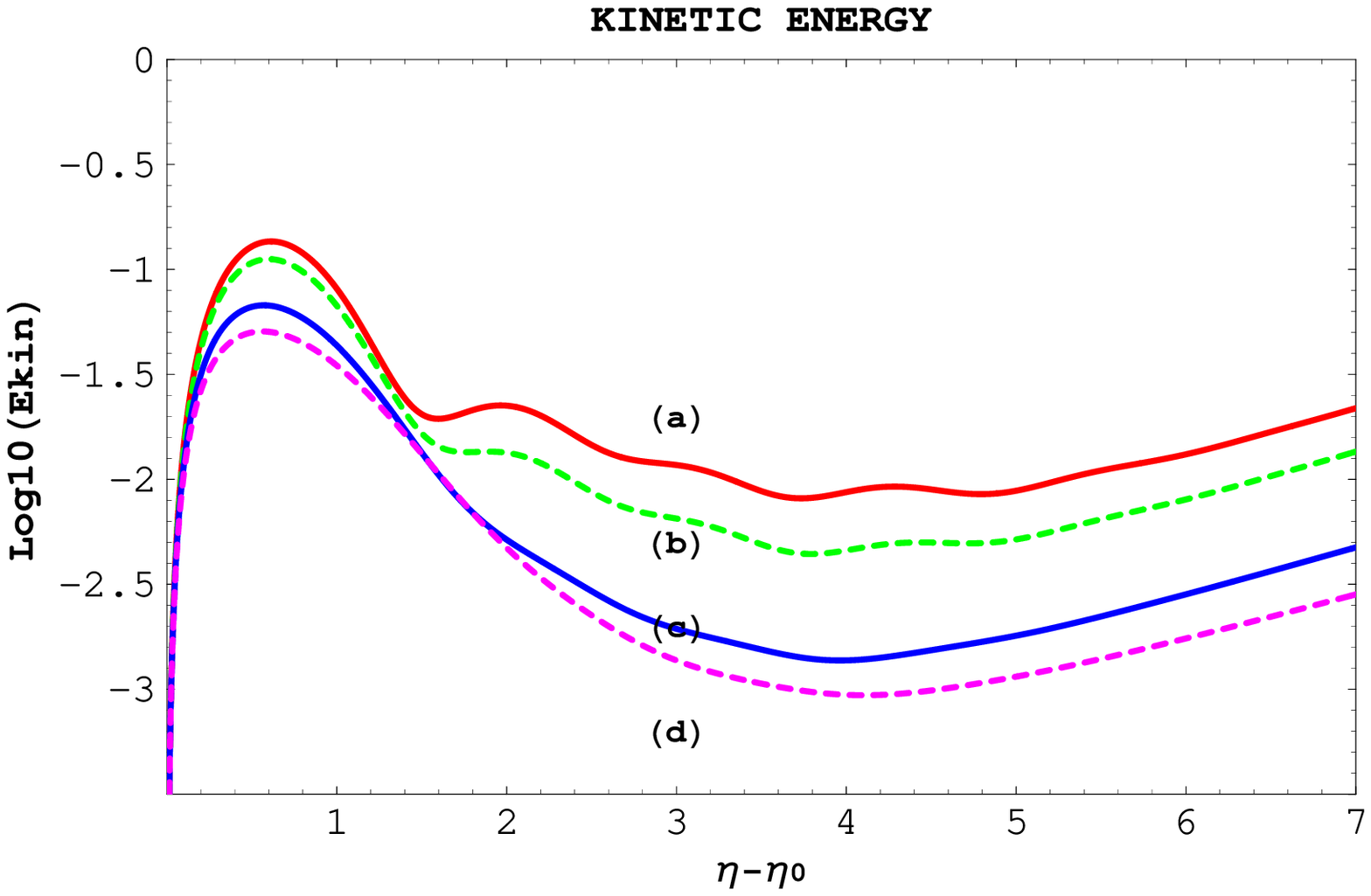} 
\includegraphics*[height=5.5cm]{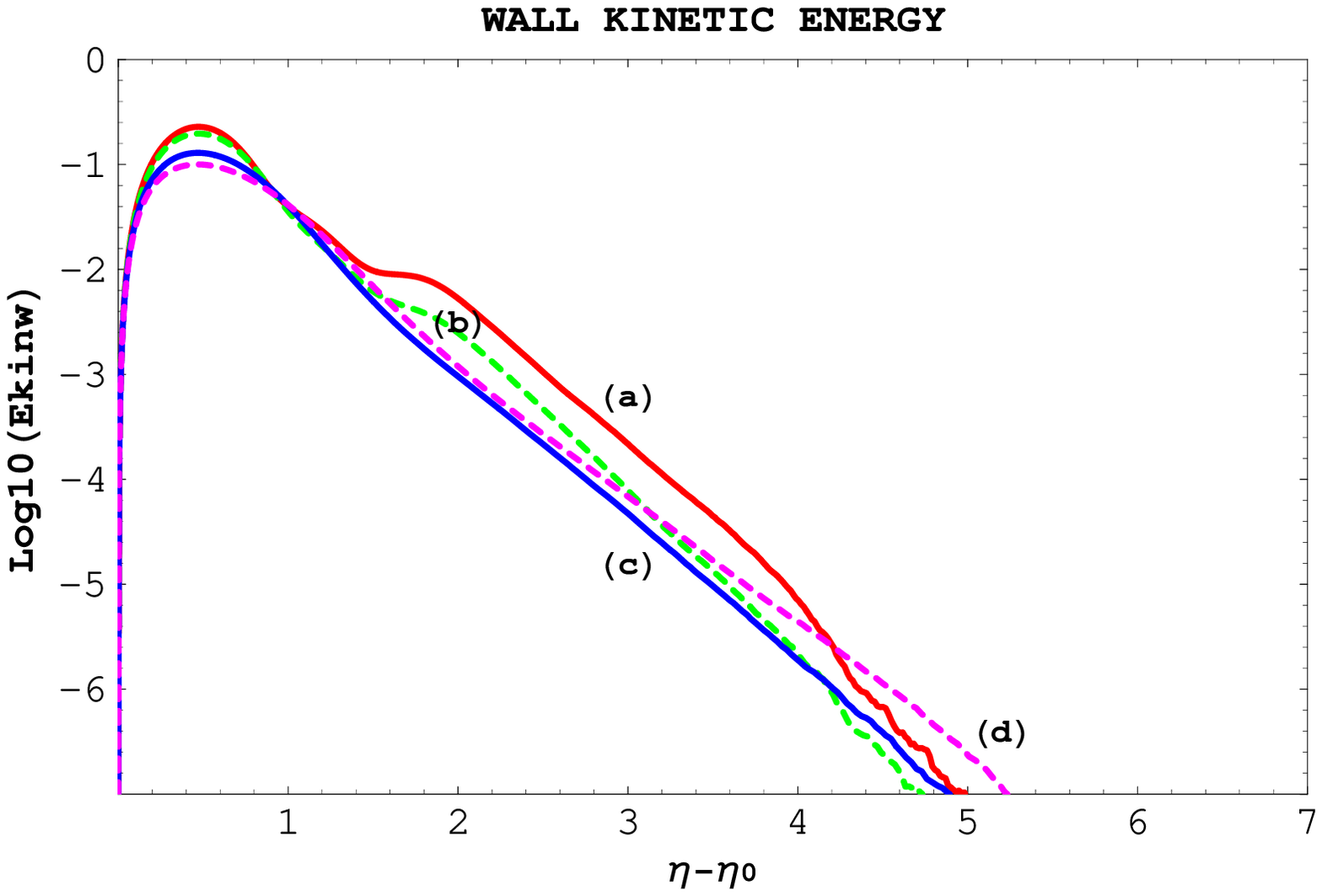} \\
\hspace*{-1.0cm}
\includegraphics*[height=5.5cm]{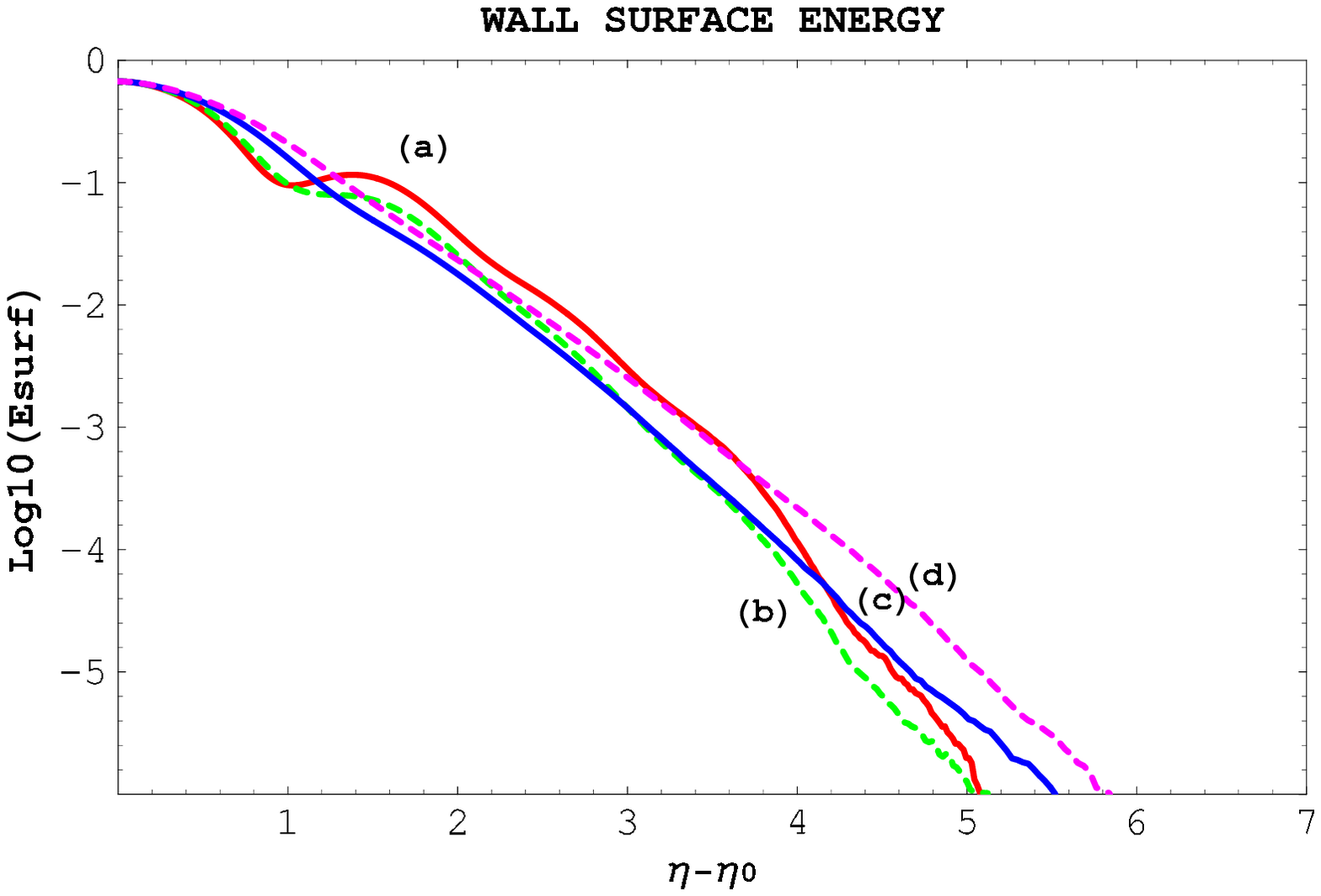}  
\includegraphics*[height=5.5cm]{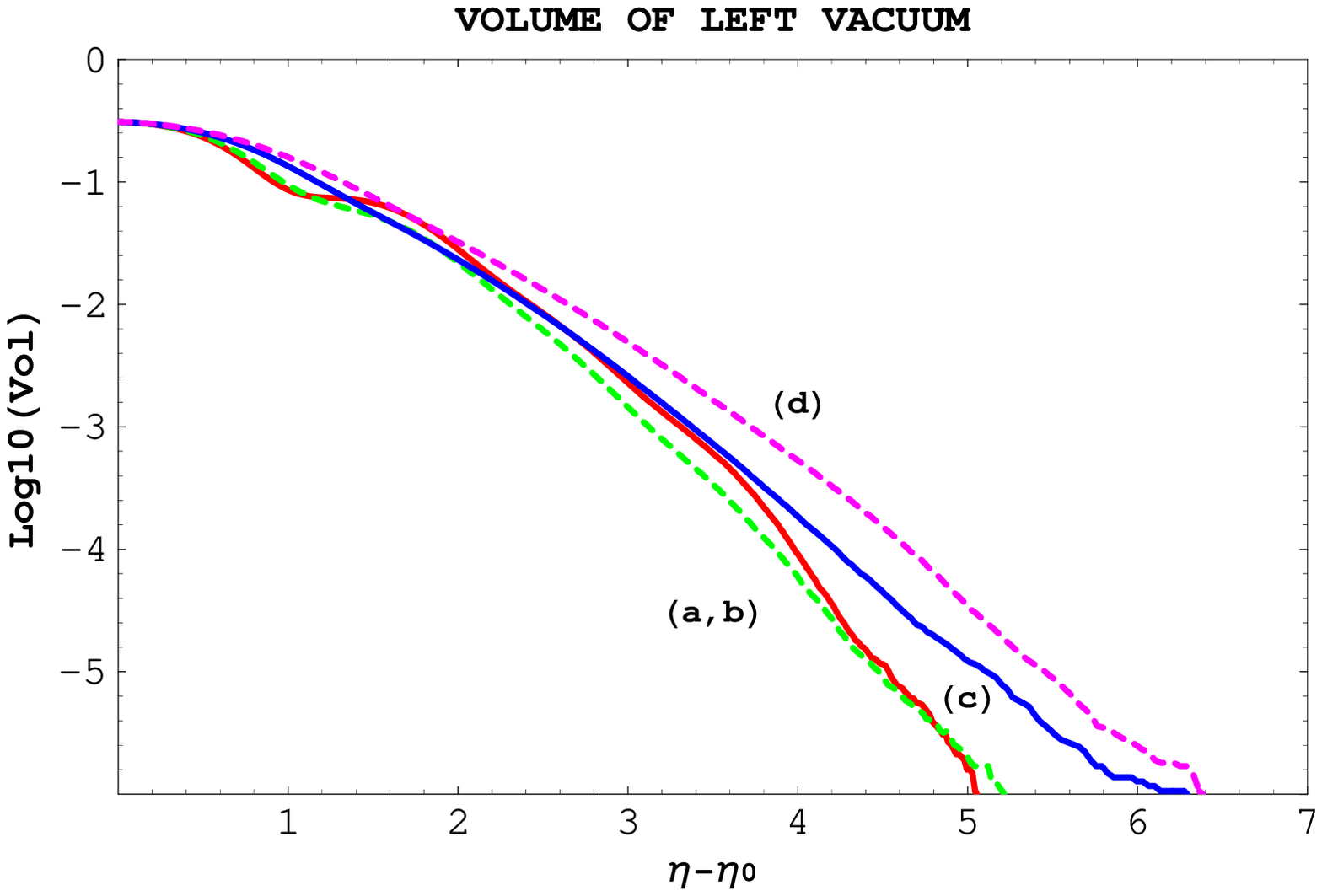}  
 
 \centering
\begin{tabular}{|c|c|c|c|c|}
  \hline
   & (a) & (b)  & (c)  &  (d)   \\
  \hline
  $\omega$ & 0.8  & 1.0  & 1.6  &  2.0   \\
  \hline
\end{tabular}
  \caption{ \it
$\log_{10}(\rm{Vol(L)})$ versus $\eta - \eta_{start}$
for different values of $\omega$.}
 \label{omega}
\end{figure}

As discussed in previous sections,
we assume that the expansion of the universe is dominated by some smooth component,
filling the universe. Then, if we go smoothly between dust
 ($\alpha=0$), and radiation, 
($\alpha=1/3$), the parameter $\omega$  changes from 2 to 1. 
We made simulations for various values of $\omega$, to see whether
this influences the evolution. The results are in Figure \ref{omega}.

All the measured observables change smoothly with $\omega$ and are not influenced
very much. In general, the faster the scale factor $a(\eta)$ 
grows, the slower the walls disappear.

\underline{Role of the horizon at the time of network creation}

\begin{figure}[!h] 
\begin{center}
\includegraphics*[height=7cm]{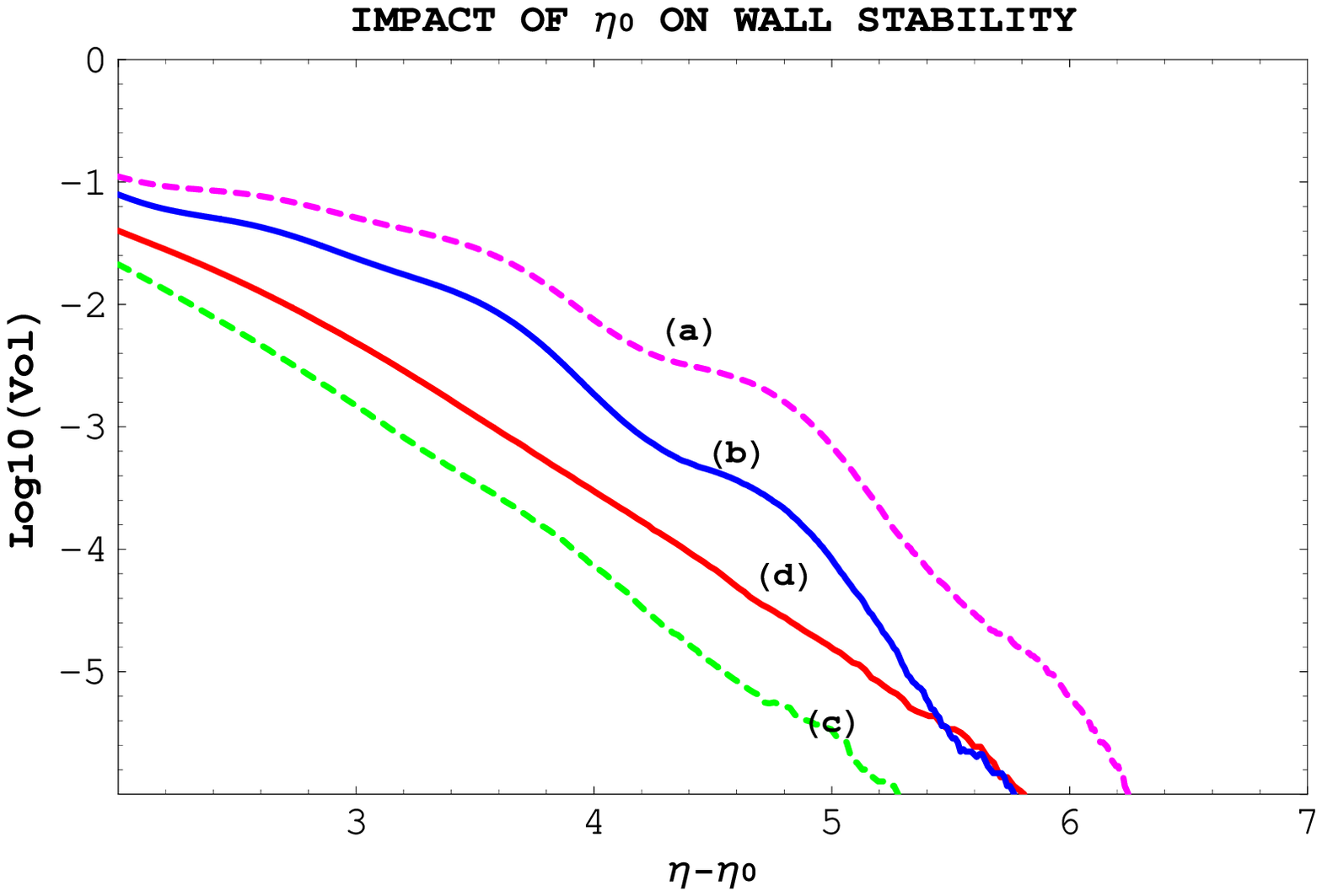} 
  \end{center}
\centering
\begin{tabular}{|c|c|c|c|c|}
\hline
 & (a) & (b) & (c) & (d) \\
  \hline
  $\eta_{start}$ & 10.0 
& 5.0  & 0.8
  & $10^{-4}$  \\
  \hline
\end{tabular}
\caption{\it $\log_{10}(\rm{Vol(L)})$ versus $\eta-\eta_{start}$ for different 
values of $\eta_0$.}
\label{etastart}
\end{figure}

If the scale factor behaves like
$\alpha(\eta) \sim \eta^\omega$, the horizon 
(inverse Hubble constant) grows with time as 
$H^{-1} = \frac{\eta^{\omega+1}}{\omega} $. 
Hence, different values of $\eta_{start}$ give different values of
the horizon at the point of the phase transition.
The results should be sensitive to that and to test how the evolution
of the field changes with the change of the initial horizon we performed
several simulations for $\eta_{start}$ changing between $10^{-4}$ and 10.
In almost all published simulations this parameter was taken
to be 1. The results are given in Figure \ref{etastart}.

Figure \ref{etastart} indicates the existence 
of two competing effects: 

(i) For very small values of $\eta_{start}$  the horizon is much
smaller than the wall width. Then, the friction term in the
equation of motion is large, and temporarily freezes the evolution 
of the network. One can see that for $\eta_{start}= 0.8$ the
network is less stable than for 
$\eta_{start}= 10^{-4}$ - in the first case the 
wall width--to--horizon ratio is smaller than in the second and 
walls evolve faster under the influence of the potential;
consequently the surface of walls decays faster. 

(ii) However, for large values of $\eta_{start}$, 
corresponding to curves (a) and (b),
many walls fit within the horizon 
and interactions between walls, of the joining and 
splitting type, become important and they tend to stabilise the network. 
This is illustrated by the fact that the network 
corresponding to $\eta_{start}=10$ is more stable that that corresponding to 
$\eta_{start}= 5$, as there is more walls inside the horizon in the first case. 

One should note that, at the initial stage of the 
evolution, in the cases where the horizon
is large, there is a period when the domain wall surface grows with time.

\subsection{Nearly-scaling solutions for  runaway potentials}

The dynamics of the walls are determined by several parameters: 
the distance to the horizon at the time the evolution starts, 
$H^{-1} = \eta^{\omega + 1} / \omega$, 
the width of the wall $\Delta$, the width of the initial didtribution $\sigma$ and the offset of the initial distribution with respect to 
the maximum of the potential $\bar{\phi} = \phi_{mean} 
- \phi_{max}$. Independently of the absolute positions of the extrema and of the absolute height of the maximum of the potential, the relations between these parameters 
shall determine the behaviour of the system. 

\underline{Width of the walls}

The domain wall width is defined as
\[
w_0 = \frac{\rm width ~of ~barrier}{\sqrt{\rm hight ~of ~barrier}}
\]
In our simulations, the width of the barrier is numerically constant and the width of
the walls will change by changing the hight of the barrier.
We have performed simulations, to check how the width of the walls influences their
evolution, for a range of $w_0$ between 0.05 and 20.0
lattice sites.

These indicate that, if the walls are thin, 
their evolution is dominated by the potential
energy, and field gradients are almost unimportant (in each site of the
lattice the field evolves independendly of the other site,
so, effectively, the walls become frozen in). In this 
case the overall behaviour of the field, as measured 
by the time dependence ot 
the mean value, is rather classical. 
The wider the walls, the milder the influence of the 
potential; gradients contribute significantly to the dynamics, and 
the evolution 
of the mean value of the field becomes non-classical.
If the walls are wide, between  2 and 20 lattice sites,
the evolution is insensitive to the domain wall width.

\underline{Initial width of the distribution}

In the runs presented in this paper
the field has been initialised randomly, according to a gaussian distribution.
If the width of the distribution is large, then also 
the probability to create many walls is larger. 
Hence, if we initialise according to a wider 
distribution the network lasts longer, meaning, that for
wider distribution, it matters less where is the 
center of the distribution (the biasing becomes less
important, since the field can climb over the barrier with a higher probability).
This is particularly significant for the run-away potential, which is asymmetric,
since the potential force (derivative) to the left of the barrier
is larger than the one to the right. Hence, the result is that, at
some stage, the false vaccum (the finite one) starts growing, because 
the force towards the left vacuum
is somewhat larger.

\underline{Initial Mean Value of the Field}

We have performed simulations for various positions 
of the center of the distribution for the runaway 
potential. If the field starts to the left of the maximum, 
even high above the barrier, then, very often,
the whole space finishes in the false vacuum. The 
reason is the asymmetry of the force in this potential,
together with the damping due to the Hubble 
expansion (and the fact that the friction 
term is proportional to the time derivative 
of the field, which may be large in such situations).
Most interesting effects are seen when the initial value of the
field is close to the maximum - then we get plenty 
of walls which disappear rather slowly.
If we want to obtain a stable network, then we have 
to start slightly to the right of the maximum,
again because of the asymmetry of the force. In these 
cases the networks exhibit nearly-scaling behaviour. 

\underline{Examples of nearly-scaling networks}

\begin{figure}[!h]
\hspace*{2.0cm} 
\vspace*{0.5 cm}
\includegraphics*[height=6cm]{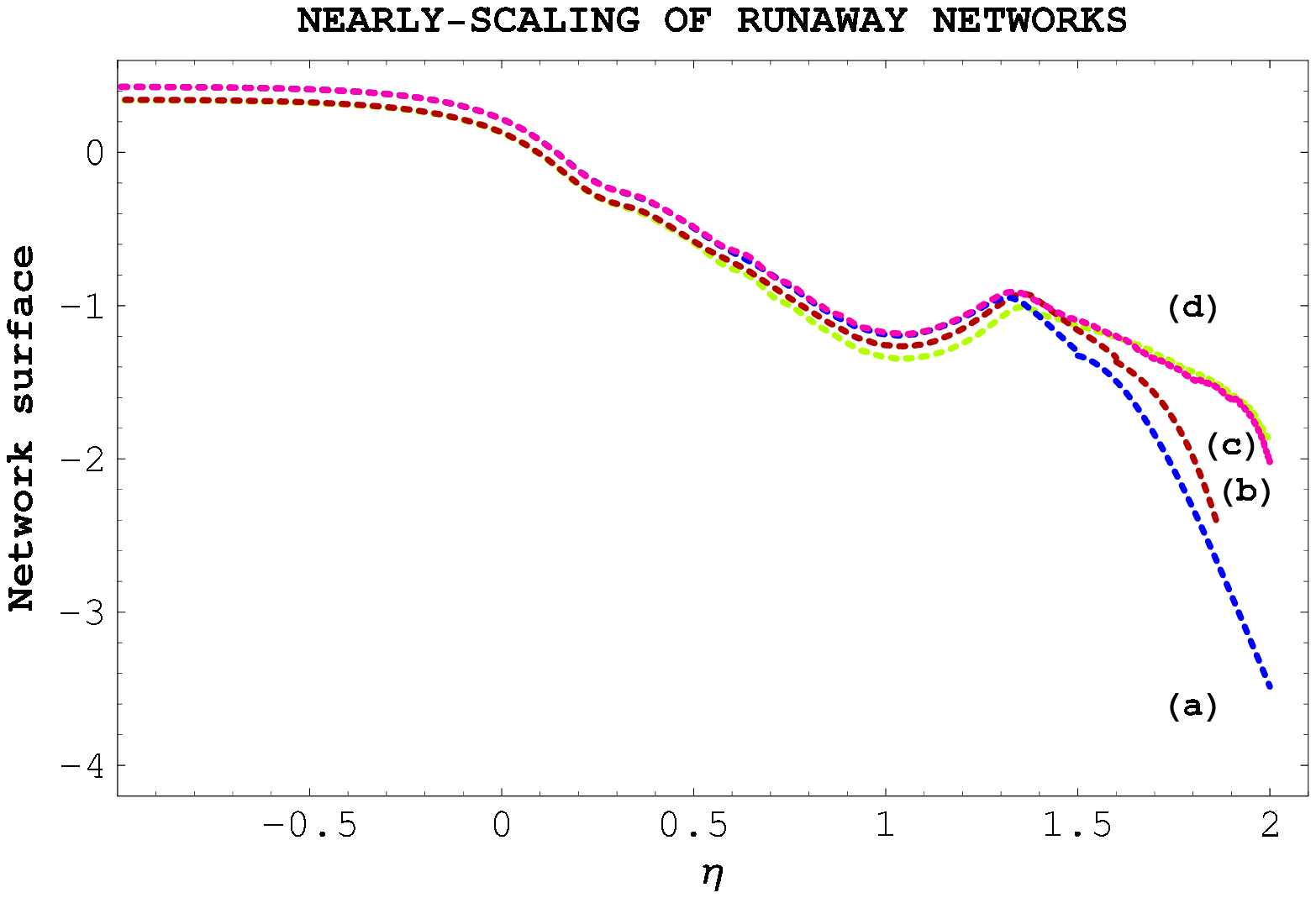} \\
\hspace*{2.0cm} 
\includegraphics*[height=6cm]{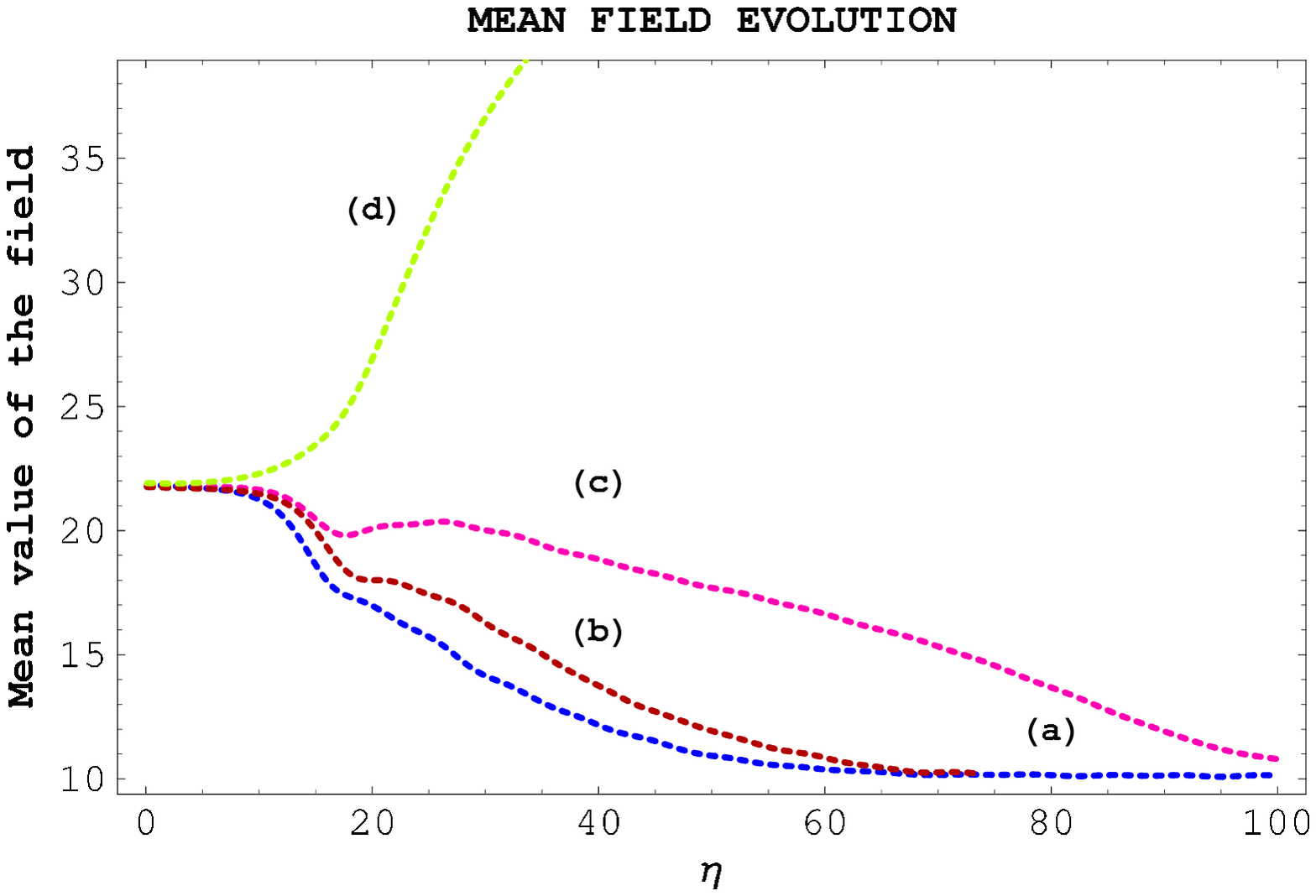}
\caption{\it Wall network evolution in runaway potentials}
\label{nstrun} 
\end{figure}

The advertised behaviour has been illustrated in Figure \ref{nstrun}.
No splitting term has been switched on in this case. 

(i) Here, the curves (a) and (b) correspond to initial 
distributions positioned at the top of the barrier 
and different widths (1.5 domain wall 
width for curve (a) and a single domain wall width for (b)).
In both cases the field evolves towards the left vaccuum, 
with the difference in the widths  playing a minor role.    \\
(ii) Curves (c) and (d) correspond to the initial mean value of 
the distribution, shifted by one twentieth 
of the wall width to the right of 
the top of the potential and width 
of the distribution equal 
to 1.5 domain wall width for (c), and to
the domain wall width for (d).

In both cases one observes a scaling behaviour of the 
network surface, however in the case of a  wider initial distribution, 
the mean field  behaves nonclassically and returns to 
the left vaccuum, so eventually the left vacuum prevails 
over the run-away behaviour.

To summarize, 
the asymmetry of the potential and its derivatives 
with respect to its maximum makes the evolution of broad and biased
distributions non-classical. 
In the limit of vanishing width of the distribution 
the classical behaviour is recovered, which is a good check of the numerical routine. 
A wide class of initial distributions leads to 
relatively short period of the existence of domain-wall 
networks, which however rather quickly disappear 
leaving behind system rolling towards infinity. This is a 
version of the Steinhardt--Brustein effect.
However,  a larger width of the distributions slowns down the decay of 
the islands of `finite' vacua. As in the case of the symmetric 
double well potential, the less favoured vacuum assumes 
the topology of compact clusters submerged 
in the sea of the run-away vacuum. The formation 
of a pseudo-infinite cluster requires a higher degree of 
fine-tuning than in the case of the double well potential.  
An important factor is the ratio of the distance to 
the horizon and the domain wall width at the time when 
the initial conditions are imposed; if this ratio is 
truly small, the disappearance of the walls becomes slower. 
This is more or less expected, as in this case the cosmic friction term 
is able to compete efficiently with the potential force. 
In the simulations shown in the figures we were assuming 
the initial horizon to be $H^{-1}_i = 10^{-4}$, which 
corresponds to walls much wider than the initial horizon. 
The small  ratio discussed above can be obtained by making 
the phase transition occur shortly before the end of inflation, 
as discussed earlier in this 
paper.    

\section{Conclusions}

The formation of large scale structure in the universe is at present one of the most 
important areas where particle physics meets cosmology. 
In particular, important contributions to structure formation 
may come from phase transitions,
especially such, when the mass order parameter is so small, 
that the characteristic scale
$1/m$ corresponds to the range of scales relevant for cosmological observations.
It is also possible that the mass of the order parameters lies in the electroweak range,
but the phase transition could be seen via its 
indirect effects. This is the case for the transition associated to supersymmetry
breaking. In the present work, using numerical and analytical
methods, we studied the physics of the domain walls that
appear during such phase transitions.
In particular, we have investigated domain wall networks and their evolution, for two
types of potential, and two ways of modeling the initial conditions after the
phase transition.
One of the potentials is the well-known 
double well potential, and the second one is the
exponential potential, characteristic of supersymmetry breaking 
via gaugino condenstation.
In both cases, we checked the evolution of domain walls as a 
function of the parameters of the 
potential, in particular (i) as a function of the non-degeneracy 
of the available vacua, and also,
(ii) as a function of a difference of probabilities of filling these vacua. To this end,
we have constructed a C++ computing code, which we used to confirm earlier results
and to extend our considerations to the new cases.

The program has been optimised, and we have found a 
theoretical estimate for the accuracy of the
integration procedure. The latter is constantly  
monitored in order to enable the 
use of an adaptive time step that greatly increases the speed of the
code while maintaining high sensitivity.
Moreover, we investigated the role of the modifications of the equation of
motion used to model the evolution on the grid. 

The simulations show compensation effects between the non-degeneracy of 
the vacua and the  asymmetry of the
probability distribution: these competing effects 
may cancel each other, resulting in the creation of slowly disappearing
metastable domain wall networks, in very general and physically interesting situations.

Extensions to other types of potentials, as 
well as a detailed study of structure formation 
within this framework shall be addressed in a separate publication.

\vspace*{0.4 cm}
{\bf Appendix I: Discretisation of equation of motion}

In order to treat the equations of motion of the scalar field and the
domain wall network numerically, we devide the universe into balls of radius 
$\ell$ much larger than $H^{-1}$,
thus covering many Hubble horizons. We will then simulate 
the evolution of the network
in a single ball of radius  $L$. The lattice site at the beginning of the simulation 
corresponds to a single Hubble horizon. Moreover, we introduce 
multi-torus topology of the grid (namely periodic boundary conditions).
To discretise the relevant equations, we 
use the ``staggered leapfrog'' method
for the second order time derivatives, and the
 Crank--Nicholson scheme for space-like derivatives. 
This means that we have second order accuracy in differentials
with respect to time and space.
The discretised equations are as follows:

$$
    \delta = \frac{1}{2} \alpha \frac{\Delta\eta}{\eta}
    \frac{d\ln{a}}{d\ln{\eta}},
$$
$$
    (\nabla^2 \phi)_{i,j,k} \equiv \phi_{i-1,j,k} + \phi_{i+1,j,k}
    + \phi_{i,j-1,k} + \phi_{i,j+1,k} + \phi_{i,j,k-1} +
    \phi_{i,j,k+1} - 6\phi_{i,j,k},
$$
\begin{eqnarray}
\label{r_roznicowe}
    \dot{\phi}^{n+\frac{1}{2}}_{i,j,k} =
    \frac{(1-\delta)\dot{\phi}^{n-\frac{1}{2}}_{i,j,k}
    + \Delta\eta \left( \nabla^2 \phi^{n}_{i,j,k} -
    a^{\beta}\frac{\partial V}{\partial \phi^{n}_{i,j,k}}
    \right)}{1+\delta},
\end{eqnarray}
$$
    \phi^{n+1}_{i,j,k} = \phi^{n}_{i,j,k}
    + \Delta \eta \dot{\phi}^{n+\frac{1}{2}}_{i,j,k}
$$
where $\eta = \eta_0 + m\Delta \eta$ (in the above,
upper indices denote time steps and lower ones
coordinates x,y and z).
For clarity 
\begin{eqnarray} \nonumber
    \phi^n_{i,j,k} \equiv \phi \left(\eta', x',y', z'\right), \\ \nonumber
    \dot{\phi}^{n+\frac{1}{2}}_{i,j,k} \equiv
    \frac{\partial \phi}{\partial \eta}\left(\eta'', x', y', z' \right), \\ \nonumber
    \frac{\partial V}{\partial \phi^n_{i,j,k}} \equiv
    \frac{\partial V}{\partial \phi} \left(\phi^n_{i,j,k}\right),
\end{eqnarray}
\begin{eqnarray} \nonumber
    x' = x_0 + i,\\ \nonumber
    y' = y_0 + j,\\ \nonumber
    z' = z_0 + k,\\ \nonumber
    \eta' = \eta_0 + n\Delta\eta,\\ \nonumber
    \eta'' = \eta_0 + (n+\frac{1}{2})\Delta\eta. \nonumber
\end{eqnarray}

Here, $\dot{\phi} \equiv \frac{\partial
\phi}{\partial \eta}$, and,  $x_0 = y_0 = z_0 = 0$.

For the mean value and the dispersion of the field, we have the following equations:

${VOL}= {N_x \cdot N_y\cdot N_z}$,

$$ \langle \phi \rangle=\sum_{i,j,k} \frac{\phi(i,j,k)}{N_x N_y N_z}
= \frac{\sum \phi}{{VOL}}.
$$

$$
\sigma_{\phi}^{2} = \langle (\phi - \langle\phi\rangle)^{2}\rangle =
\langle\phi^2\rangle - \langle\phi\rangle^2.
$$

$$
\frac{\sigma_{\phi}}{\langle\phi\rangle} = \frac{\sqrt{\frac{\sum
\phi^2}{VOL}-\langle\phi\rangle^2}}{{\langle\phi\rangle}}, ~~
\langle\phi^2\rangle = \frac{\sum \phi(i,j,k)^2}{VOL}.$$

The kinetic energy is given by

$$
E_{kin} = \frac{\sum \dot{\phi^2}}{VOL}.
$$
In all simulations we assume that the field was initially 
at rest $(\Phi^{\prime} =0)$, while, for the surface energy we have:

\begin{eqnarray}
 A= \int\vec{n} \cdot \vec{dA} = \Delta A \sum_{laczniki}
    \frac{\delta^{\pm}}{|\cos{\theta_x}| + |\cos{\theta_y}| + |\cos{\theta_z}|} = \\
    = \Delta A \sum_{laczniki} \frac{\delta^{\pm}}{|n_x| + |n_y| + |n_z|} = \\
    = \Delta A \sum_{laczniki} \frac{|\nabla\phi|}
    {|\frac{\partial \phi}{\partial x}| + |\frac{\partial \phi}{\partial y}|
    + |\frac{\partial \phi}{\partial z}|}\; ,
\end{eqnarray}

\begin{equation}
\delta^{\pm} = \left\{
\begin{array}{ll}
    1, & \hbox{if a link crosses a wall}, \\
    0, & \hbox{if it does not .} \\
\end{array} \right.
\end{equation}

$ E_{\rm{surf}} = \sigma A $, where $\sigma$ is the tension and $A$ the surface.
In what follows, we take $\sigma=1$ and 
$$ E_{\rm{surf}} = \frac{A}{VOL}. $$

The kinetic energy of the walls is given by

$$
E_{\rm{kw}} = \sum_{links} \frac{1}{2} \left[
\dot{\phi}({\overrightarrow{x}})^2 + \dot{\phi}({\overrightarrow{x}
+ \overrightarrow{n}})^2 \right].
$$
(with $\dot{\phi}$ computed at the position of the wall).
To calculate the volume of the vacuum,
we normalise to the total volume, namely take the
 number of left-vacua and right vacua over the total
number of vacua.

In our code, instead of looking for the 
extrema of the potential analytically,
we use numerical methods which are simpler (particularly for
runaway potentials):
$$ \frac{dV}{d\phi} \left(\phi \right) = \frac{V(\phi + \epsilon) -
V(\phi - \epsilon)}{2 \epsilon},
$$ and we take $\epsilon = 10^{-4} $. The accuracy is 
 $\epsilon^2 \frac{d^2V}{d\phi^2}$.

\vspace*{0.4 cm}
{\bf Appendix II: 
Technical discussion about optimisation of size of time step}

The basic parameter that determines the accuracy of the simulation is the
time step, which must be small, so that the discretised equation must well-represent
the continues one. However, the time step cannot be too small, because there are many
steps in the integration and numerical mistakes cummulate (each step introduces an
error).

We have seen that the field changes rapidly at the beginning of the simulation,
which is due to the random, non-equilibium initialisation and somewhat later, by
the fact that domain walls get created rapidly and then interact very often (since there
are many of them). After some time, the field changes at a slower rate, and its configuration
becomes more regular and more stable. Consequently, we should 
change at some point the time step of the simulation 
(smaller one at the beginning, when the evolution is rapid, 
and larger, at later times).

The change of the field depends on the time step, and on the value of the time
derivative in the next integration step
$$
   \Delta \phi = \Delta \eta \dot{\phi}.
$$
Now, let's look at the evolution of the time derivative.
The change of the time derivative over the time $\delta \eta$ is
$$
    \Delta \dot{\phi} \equiv \dot{\phi} -
    \frac{(1-\delta)\dot{\phi} + \Delta\eta \left( \nabla^2 \phi -
    a^\beta \frac{\partial V}{\partial \phi} \right)}{1+\delta}, \\
$$
which gives
$$
    \Delta \dot{\phi} = \frac{1}{{1+\delta}} \left[
    -2\delta \dot{\phi} + \Delta\eta\ \left( \nabla^2 \phi - a^{\beta}V' \right) \right]. \\
$$
The $\delta$ is given by 
$$
      \delta = \frac{1}{2} \alpha \frac{\Delta\eta}{\eta} \frac{d\ln{a}}{d\ln{\eta}}.\\
$$
Since $\alpha \sim 2$, and 
$\frac{d\ln{a}}{d\ln{\eta}}\sim 1,$ 
$$
      \delta \sim \frac{\Delta\eta}{\eta}, so
$$
$$
      \frac{1}{1+\delta} \sim 1 - \frac{\Delta\eta}{\eta}.
$$
Substituting, we get
$$
    \Delta \dot{\phi} \sim  \Delta \eta  \left[
    -2\frac{\dot{\phi}}{\eta} + \nabla^2 \phi - a^{\beta} V'\right]
    \left( 1- \frac{\Delta\eta}{\eta} \right).
$$

The expression in the square bracket can be estimated from above, by its largest
value, taken anywhere in the lattice.

\begin{equation}
    |\Delta \dot{\phi}| \le \Delta \eta \left(
    \frac{-2|\dot{\phi}|_{max}}{\eta} + |\nabla^2 \phi|_{max} + |a^{\beta} V'|_{max}
    \right).
\end{equation}

Now, we can follow several strategies;

(i) demand that the change to the field is as small as possible
$$
    \frac{\Delta \phi}{\phi} \ll 1,
$$

(ii) demand that the average change of the field 
on the whole lattice should be smaller than 1
$$
    \left< \frac{\Delta \phi}{\phi} \right> \ll 1,
$$

(iii) request that the maximal change of the field with 
respect to the wall width is much smaller than 1
$$
    \frac{\Delta \phi}{w_0} \ll 1.
$$

In our simulation, we have followed the last path, thus,
$$
    \frac{\Delta \phi}{w_0} \le \kappa
$$
 where $\kappa$
is the requested accuracy. From this inequality, it turns out that
$$
\Delta \eta = \sqrt{ \frac{\kappa
w_0}{\frac{-2|\dot{\phi}|_{max}}{\eta} + |\nabla^2 \phi|_{max} +
|a^{\beta} V'|_{max}}}.
$$
This is the estimated optimal time step.

To fix the optimal accuracy, we have performed 
several simulations, looking for the change of the
results with respect to changes of $\kappa$. 
We used a 2d lattice with a size of 248 $\times$ 248, for 
$w_0 = 356$. We also made a simulation on larger lattice,
3072 $\times$  3072, to understand the dependence 
on the lattice size. The results appear in Table 2 below.

\begin{table}[!h]
 \centering

\begin{tabular}{|c|c|c|c|c|c|c|}
  \hline
  $\kappa$             & $10^{-1}$ & $10^{-2}$ & $10^{-3}$ & $10^{-4}$ & $10^{-5}$& $10^{-6}$ \\
  \hline
  $({\Delta \eta})_{min}$  & $1.85\cdot 10^{-1}$ & $5.89\cdot 10^{-2}$   & $1.79\cdot 10^{-2}$  & $5.72\cdot 10^{-3}$  & $1.85\cdot 10^{-3}$ & $5.87\cdot 10^{-4}$ \\
  $({\Delta \eta})_{max}$  & $6.16\cdot 10^{-1}$& $2.03\cdot 10^{-1}$    & $6.41\cdot 10^{-2}$  & $2.02\cdot 10^{-2}$  & $6.45\cdot 10^{-3}$ & $2.09\cdot 10^{-3}$ \\
  \hline
\end{tabular}
\label{tabb2}
\caption{\it 
  $({\Delta \eta})_{min}$  versus $\kappa$.}
\end{table}

It turns out that the kinetic energy, 
mean value of field and dispersion
of the field are not sensitive to the time-step, however, the surface energy,
surface of walls and volume of different vacua, are much more dependent on it
(in fact, the most sensitive statistical observable is the ratio of the surface 
of the walls over the volume of the subdominant vacuum - namely the 
rate of change of the volume of the walls).

In the simulations described so far in the literature,
there is no analysis of the accuracy of the
results, and the time-step is usually not given. Instead,
large series of simulations are performed and the results are
averaged. However, it may be that the numerical noise is being
averaged as well. The new element arising from our simulation is
that, with appropriately chosen time-step, the plots of the parameters
are simular for different runs, which makes them more reliable.

In our procedure, we 
first chose the optimal time step for a given accuracy, $\kappa$,
and then we make small series of simulations (requiring among others
less computing power).

The size of the lattice we use is large, about 4 millions of horizons, thus, 
with the optimal time step, the observables are computed very precisely.
The point of time when the graphs stop being smooth (have discontinuities) is
 interpreted as the point where the accuracy is lost.
From then onwards, the predictions for the
observables can be treated only qualitatively.
After careful analysis, we have fixed $\kappa$ to be $ 10^{-4}$
 which gives us the time step to be in the range 0.005 and 0.02.

\vspace*{0.4 cm}
{\bf Appendix III: Size of the Lattice}

Representing the field on a large lattice requires a lot of memory.
A single simulation in 2D on a lattice of 2048 $\times$
2048 with $\kappa = 10^{-4}$ 
requires more than 10 hours of computing time, to reach $\eta_{stop} \sim 100$.

The resolution $\delta$ (precision) that we previously discussed,
is the smallest visible relative change of the field 
statistics - we estimated this by
looking at the plots of surface energy of 
the walls comparing them to the simulation performed
on the largest possible lattice. 
The results are given in Table 3, for 2d and 3d cases.

\begin{table}[!h]
 \centering
\begin{tabular}{|c|c|c|c|c|c|c|}
  \hline
  $n_x$  &   786     & 1024              & 1536      & 2048             & 3072            & 4096 \\
  \hline
  $\delta$ & $10^{-3}$ & $2\cdot 10^{-4}$  & $10^{-4}$ & $5\cdot 10^{-5}$ & $1\cdot 10^{-5}$& ${2\cdot 10^{-6}}$ \\
  \hline
\end{tabular}

\vspace*{0.5 cm}
\begin{tabular}{|c|c|c|c|c|c|c|}
  \hline
  $n_x$ &   80              & 100               
& 128       & 160              & 200               & 256 \\
  \hline
  $\delta$ & $7\cdot 10^{-5}$  & $4\cdot 10^{-5}$  
& $10^{-5}$ & $5\cdot 10^{-5}$ & $2.5\cdot 10^{-6}$& ${6.2\cdot 10^{-7}}$ \\
  \hline
\end{tabular}
\label{tabb3}
\caption{\it Resolution versus lattice size in 2d (upper panel) and in 3d simulations.}
\end{table}

We have found that the logarithm of the resolution $\delta$  
is linear in the
lattice size, and the best fit for 2d is given by 

$$
\log_{10}{\delta} = -3.237 - 1.88 \cdot 10^{3}\cdot {n_x}^{3/2},
~~\delta \sim 10^{-3.237} \cdot 10^{-\frac{{n_x}^{3/2}}{1526}}.
$$

For 3d the logarithm of the resolution 
is proportional to the power of $n$:

$$
\log_{10}{\delta} = -3 - 6.55 \cdot 10^{-4}\cdot n_x,
~~
\delta \sim 10^{-3} \cdot 10^{-\frac{n_x}{1526}}.
$$

In our simulation, we need these formulas to judge when the number 
of domain walls is too small to trust the numerical results.
Technically, because
we are using periodic boundary conditions, after a finite time
(which is
of the order of the lattice size over the velocity of the wall),  
a wall which leaves the
horizon could return, because it comes to the lattice 
from the other side.
This sets the
limit of the simulations. Thus, looking at the simulations, we can see that the average
velocity of the walls is  0.5, which means that the return time is approximately equal 2 times
the lattice size. Thus, this is the absolute upper limit on the huseful range of $\eta$s
- because here we have a large lattice with periodic boundary conditions.

We have verified that for the lattices we have used, the role of the periodic boundary conditions
is negligible.

\vspace*{0.4 cm}
{\bf Appendix IV: Monitoring of the position of the Minima of different 
Potentials}

The position of the extrema of the potential is monitored both analytically
and numerically.

\underline{Double Well  Potential}

For the potential

$$V(\phi) \rightarrow V(\phi) - \epsilon V_0 \phi$$

with

$$ V(\phi) = V_0 \left( \left(\frac{\phi}{\phi_0}\right)^2 -1\right)^2. $$

The Taylor expansion of the potential is

$$
V(\phi) = V_0 + \frac{1}{2} \left(\frac{V_0}{v^2} \right)\phi^2 +
\frac{1}{4!} \left(\frac{V_0}{v^4} \right)\phi^4 + \dots
$$

and, for a small non-degeneracy parameter $\epsilon$ we can easily find 
the position for the extrema of the potential:

(maximum)

$$ \phi \ = \ - \frac{1}{4{\phi_0}^2}\ \epsilon - \frac{1}{64{\phi_0}^8}\ \epsilon^3
        \ - \frac{3}{1024{\phi_0}^14}\ \epsilon^5 \ - \ \frac{3}{4096{\phi_0}^20}\ \epsilon^7 \ldots $$

(minimum)

\begin{eqnarray}
\nonumber \phi =  -\phi_0 + \frac{1}{8{\phi_0}^2}\ \epsilon -
\frac{1}{128{\phi_0}^5}\ \epsilon^2
  + \frac{1}{128{\phi_0}^8}\ \epsilon^3 - \frac{105}{32768{\phi_0}^{11}}\ \epsilon^4 + \\ \nonumber
  + \frac{3}{2048{\phi_0}^{14}}\ \epsilon^5  - \frac{3003}{4194304{\phi_0}^{17}}\ \epsilon^6
  + \frac{3}{8192{\phi_0}^{20}}\ \epsilon^7 \ldots, \\ \nonumber\\
\nonumber \phi =  \phi_0 + \frac{1}{8{\phi_0}^2}\ \epsilon +
\frac{1}{128{\phi_0}^5}\ \epsilon^2
  + \frac{1}{128{\phi_0}^8}\ \epsilon^3 + \frac{105}{32768{\phi_0}^{11}}\ \epsilon^4 + \\ \nonumber
  + \frac{3}{2048{\phi_0}^{14}}\ \epsilon^5  + \frac{3003}{4194304{\phi_0}^{17}}\ \epsilon^6
  + \frac{3}{8192{\phi_0}^{20}}\ \epsilon^7 \ldots.
\end{eqnarray}

The important thing is the position of the maximum, because this we use as the
border between the left and right vacuum. One can always find numerically the position
of the extrema, via the 
Newton--Raphson method: $x_{i+1} = x_i
-\frac{f(x_i)}{f'(x_i)}$ .

\underline{Runaway Potential}

For

$$ V(s) = \frac{1}{2s} {\left( A(2s+N_1) e^{-\frac{s}{N_1}} - B(2s+N_2) e^{-\frac{s}{N_2}}\right)}^2,$$

(where $s$ 
is canonically normalised), we find the extrema as follows:

\begin{eqnarray}
 \nonumber
   \frac{dV}{ds} =
 - \frac{1}{2 N_1 N_2 s^2} e^{-2 \frac{s(N_1 + N_2)}{N_1 N_2}}
   \left[ A(2s+N_1) e^{\frac{s}{N_2}} - B(2s+N_2) e^{\frac{s}{N_1}} \right] \cdot \\  \nonumber
   \left[ A N_2 (N_1^2 + 4s^2) e^{\frac{s}{N_2}}  -  B N_1 (N_2^2 + 4s^2) e^{\frac{s}{N_1}}
   \right].
\end{eqnarray}

Here, we have two brackets and either the one or the other vanishes. This gives
the following two conditions:

$$ e^{s \frac{N_1 - N_2}{N_1 N_2}} = \frac{B(N_2 + 2s)}{A(N_1 + 2s)}, $$

$$ e^{s \frac{N_1 - N_2}{N_1 N_2}} = \frac{B N_1 (N_2^2 + 4s^2)}{A N_2 (N_1^2 + 4s^2)}.$$

These two conditions are non-linear and cannot be solved algebraically, but can be solved
iterativelly, step-by-step.

$$ s_{(i+1)} = \frac{N_1 N_2}{ N_1 - N_2} \ln 
\left[ \frac{B(N_2 + 2s_{(i)})}{A(N_1 + 2s_{(i)})} \right],$$
$$ s_{(i+1)} = \frac{N_1 N_2}{ N_1 - N_2} 
\ln \left[ \frac{B N_1 (N_2^2 + 4s^2_{(i)})}{A N_2 (N_1^2 + 4s^2_{(i)})} \right],$$

$$s_{(0)} = \frac{N_1 N_2}{ N_1 - N_2}.$$

To remove the degeneracy of the vacua, we add a term that looks like
 $\sim \frac{\epsilon}{s^2}$.

\vspace*{0.4 cm}
{\bf Acknowledgements}
This work was partially supported by the EC 6th Framework
Programmes MRTN-CT-2006-035863 and MRTN-CT-2004-503369, 
the  grant MNiSW  N202 176 31/3844 and the TOK Project  
MTKD-CT-2005-029466. The research of S. Lola is co-funded by 
the FP6 Marie  Curie Excellence Grant MEXT-CT-2004-014297.  
Z. Lalak and S. Lola would like to thank 
the Theory Division of CERN for kind hospitality during a 
large part of this work.

\end{document}